\documentclass[prd,twocolumn,
amsmath,
amssymb,
aps,
preprintnumbers,nofootinbib]{revtex4-1}
\pdfoutput=1
\usepackage{graphicx}% Include figure files
\usepackage{dcolumn}% Align table columns on decimal point
\usepackage{bm}% bold math
\usepackage{xcolor, hyperref}
\usepackage[normalem]{ulem}
\usepackage{booktabs}
\usepackage{multirow}
\usepackage[export]{adjustbox}
\usepackage{floatrow}
\newfloatcommand{capbtabbox}{table}[][3.5in]
\newfloatcommand{capbfigbox}{figure}[][2.3in]
\usepackage{blindtext}
\usepackage{booktabs}
\usepackage[normalem]{ulem}
\usepackage{bbm}
\usepackage{booktabs}      % tables
\usepackage{youngtab}

 \newcommand{\MeV}{\,\text{MeV}} \newcommand{\keV}{\,\text{keV}} \newcommand{\eV}{\,\text{eV}} \newcommand{\cm}{\,\text{cm}}

\newcommand\Tstrut{\rule{0pt}{2.5ex}}         

\newcommand{\sigmabare}{\overline{\sigma}_e}

\renewcommand{\vec}[1]{\mathbf{#1}} % vectors are boldface

\renewcommand{\eqref}[1]{Eq.~(\ref{#1})}

\DeclareMathOperator\erf{erf}

\begin{document}

\title{
Dark Matter Daily Modulation With Anisotropic Organic Crystals
}

\author{Carlos Blanco$^{a,b}$}%\note{ORCID: http://orcid.org/0000-0001-8971-834X}
\email{carlosblanco2718@princeton.edu}

\author{Yonatan Kahn$^{c,d}$}%\note{ORCID: http://orcid.org/XXXXXXX}
\email{yfkahn@illinois.edu}

\author{Benjamin Lillard$^{c,d}$}%\note{ORCID: http://orcid.org/XXXXXXX}
\email{blillard@illinois.edu}

\author{Samuel D. McDermott$^{e}$}%\note{ORCID: http://orcid.org/XXXXXXX}
\email{sammcd00@fnal.gov}

\affiliation{${}^a$Stockholm University and The Oskar Klein Centre for Cosmoparticle Physics,  Alba Nova, 10691 Stockholm, Sweden}
\affiliation{${}^b$Department of Physics, Princeton University, Princeton, NJ 08544}
\affiliation{${}^c$Department of Physics, University of Illinois at Urbana-Champaign, Urbana, IL 61801}
\affiliation{${}^d$Illinois Center for Advanced Study of the Universe, University of Illinois at Urbana-Champaign, Urbana, IL 61801}
\affiliation{${}^e$Fermi  National  Accelerator  Laboratory,  Batavia,  IL  60510  USA}
\date{\today}

\preprint{FERMILAB-PUB-21-066-T}

\begin{abstract}
Aromatic organic compounds, because of their small excitation energies $\sim \mathcal O$(few eV) and scintillating properties, are promising targets for detecting dark matter of mass $\sim \mathcal O$(few MeV). Additionally, their planar molecular structures lead to large anisotropies in the electronic wavefunctions, yielding a significant daily modulation in the event rate expected to be observed in crystals of these molecules. We characterize the daily modulation rate of dark matter interacting with an anisotropic scintillating organic crystal such as trans-stilbene, and show that daily modulation is an $\sim \mathcal O$(1) fraction of the total rate for small DM masses and comparable to, or larger than, the $\sim 10\%$ annual modulation fraction at large DM masses. As we discuss in detail, this modulation provides significant leverage for detecting or excluding dark matter scattering, even in the presence of a non-negligible background rate. Assuming a non-modulating background rate of 1/min/kg that scales with total exposure, we find that a 100${\rm kg \cdot yr}$ experiment is sensitive to the cross section corresponding to the correct relic density for dark matter masses between $1.3-14\MeV$ ($1.5-1000\MeV$) if dark matter interacts via a heavy (light) mediator. This modulation can be understood using an effective velocity scale $v^* = \Delta E/q^*$, where $\Delta E$ is the electronic transition energy and $q^*$ is a characteristic momentum scale of the electronic orbitals. We also characterize promising future directions for the development of scintillating organic crystals as dark matter detectors.
\end{abstract}
\pacs{Valid PACS appear here}
\maketitle

\section{Introduction}

Dark matter-electron scattering is a promising search strategy for sub-GeV dark matter (DM)~\cite{Essig:2011nj,Graham:2012su,Essig:2015cda,Lee:2015qva,Hochberg:2015pha,Hochberg:2015fth,Alexander:2016aln,Derenzo:2016fse,Hochberg:2016ntt,Kavanagh:2016pyr,Emken:2017erx,Emken:2017qmp,Battaglieri:2017aum,Essig:2017kqs,Cavoto:2017otc,Hochberg:2017wce,Essig:2018tss,Emken:2018run,Ema:2018bih,Geilhufe:2018gry,Baxter:2019pnz,Essig:2019xkx,Emken:2019tni,Hochberg:2019cyy,Trickle:2019nya,Griffin:2019mvc,Coskuner:2019odd,Geilhufe:2019ndy,Catena:2019gfa,Blanco:2019lrf,Kurinsky:2019pgb,Kurinsky:2020dpb,Griffin:2020lgd,Radick:2020qip,Gelmini:2020xir,Trickle:2020oki,Du:2020ldo,Hochberg:2021pkt,Knapen:2021run,Essig:2012yx,Tiffenberg:2017aac,Romani:2017iwi,Crisler:2018gci,Agnese:2018col,Agnes:2018oej,Settimo:2018qcm,Akerib:2018hck,Abramoff:2019dfb,Aguilar-Arevalo:2019wdi,Aprile:2019xxb,Barak:2020fql,Arnaud:2020svb,Amaral:2020ryn}. In molecular or solid-state systems, atoms are close enough that electronic orbitals overlap significantly, lowering the electronic excitation energies to the eV scale and thus allowing detection of DM particles with $\sim {\rm MeV}$-scale mass which carry eV-scale kinetic energy. Moreover, solid-state systems can exhibit anisotropic electronic wavefunctions (see for example \cite{Hochberg:2016ntt,Hochberg:2017wce,Coskuner:2019odd,Geilhufe:2018gry,Geilhufe:2019ndy,Griffin:2020lgd}), enabling directional detection schemes which leverage the characteristic signature of the daily modulation of the direction of the DM wind in the lab frame (first noted in the context of multiple scattering from terrestrial overburden in \cite{Collar:1992qc,Collar:1993ss,Hasenbalg:1997hs}, followed by the connection to directional detection in \cite{Avignone:2008cw}).

In this paper, we focus on, and advocate for, a particular class of detector materials for DM-electron scattering: aromatic organic crystals. These compounds have numerous advantages, both practical and theoretical, over existing detectors. Their molecular structures consist primarily of hexagonal carbon rings with alternating single and double bonds (see Fig.~\ref{fig:struct}), and the excited molecular electronic levels at $\mathcal{O}(5 \ {\rm eV})$ above the ground state can de-excite by emitting a scintillation photon with $\mathcal{O}(1)$ bulk quantum efficiency. As a subset of the authors showed in a previous paper \cite{Blanco:2019lrf}, simple organic liquids like benzene and its analogs have superior reach per unit target mass compared to single-electron threshold semiconductor detectors for DM heavier than about 10 MeV. Furthermore, as noted in \cite{Blanco:2019lrf} and as discussed at length in the present work, the planar structure of the molecules leads to a marked anisotropy in the electronic wavefunctions which is absent in silicon and germanium detectors and which allows for directional detection: even though the scintillation \emph{emission} is isotropic, the \emph{excitation rate} to the scintillation level depends strongly on the direction of the incoming DM.

Building on our previous work, in this paper we focus on larger organic molecules which are solids at room temperature with known bulk quantum efficiencies at cryogenic temperatures. Furthermore, we focus on molecules with reduced in-plane symmetry. This additional asymmetry means that the lowest-energy transitions are not suppressed, distinguishing them from simpler molecules like benzene. The crystal lattice effects in these organic crystals are small enough that the electronic structure closely resembles that of the isolated molecules. Single-crystal scintillators can have Avogadro's number of unit cells containing the same relative orientations of the molecules, allowing the anisotropic response to persist. From a practical standpoint, single-crystal samples of trans-stilbene (t-stilbene), which we focus on in this paper, can be manufactured at kilogram scale with order-1 scintillation efficiency, such that even a single scintillation photon produced from a DM scattering event has a high probability of being detected from a large-mass sample~\cite{Zaitseva2015}.

These practicalities suggest that running a t-stilbene experiment with a large exposure is feasible in the near future, so we consider possible interpretations of plausible near-future experimental data. 
%We find that the daily modulation provides a meaningful handle for extending the sensitivity of a t-stilbene detector deep into well-motivated DM parameter space. 
As most sources of background noise (such as radioactive impurities in the target material) are constant in time, 
daily modulation provides a way to detect dark matter even without reducing the background rate to zero.
% significantly enhances the sensitivity of a t-stilbene detector.
% the sensitivity of a t-stilbene detector can be extended deep into DM parameter space by searching for a modulating signal.
Cosmic rays, the primary time-varying external background, have a daily modulation that has been constrained to be below the level of 0.1\% in underground facilities~\cite{COSINE:2020jml}, so a modulating signal with a significantly larger amplitude would provide a clear detection of dark matter.
% daily modulation provides a meaningful handle for extending the sensitivity of a t-stilbene detector deep into well-motivated DM parameter space.
%Most sources of background noise, such as radioactive impurities in the target material, are constant 
% The endogenous background in such an experiment are composed primarily of radio impurities which produce rates that are constant in time. 
% The cosmic ray background is the primary external background whose daily modulation is constrained to be below the level of 0.1\% in underground facilities~\cite{COSINE-100:2020jml}. 

For example,
assuming an average observed rate of $1/60 {\rm \, Hz \, kg^{-1}}$, and without incorporating expectations for daily modulation, the assumed (constant) background rate would limit the reach to approximately an order of magnitude above the interesting parameter space, with no prospects for improvement over time or with a larger experiment. However, a 1$\,$kg$\cdot$yr exposure reaches DM relic density targets if the data are interpreted with the expected modulation information. The sensitivity of a modulation analysis continues to scale with $({\rm exposure})^{1/2}$ even without mitigating backgrounds, so a larger 100$\,$kg$\cdot$yr exposure improves the reach by an order of magnitude. This probes the relic density target for DM masses $1.3 \lesssim m_\chi \lesssim 14\MeV$ if the DM interacts through a heavy mediator, or the range $1.5 \lesssim m_\chi \lesssim 1000\MeV$ if the DM achieves its relic abundance through freeze-in via a light kinetically-mixed dark photon. Solid-state organic scintillator detectors would therefore greatly reduce the necessity for a low-threshold zero-background experiment in order to conclusively discover or exclude DM.

As a consequence of our analysis of organic crystals, we point out a generic feature of DM-electron scattering in condensed matter systems: daily modulation is governed by the relationship between the DM velocity and an effective electron velocity $v^* \equiv \Delta E/q^*$, where $\Delta E$ is the energy of an electronic transition and $q^*$ is the typical momentum scale for the electron wavefunctions governing the transition. A necessary condition for daily modulation is anisotropy of the molecular form factor for electronic transitions, but for the anisotropy to be kinematically accessible, the transition needs to be either near a kinematic threshold (for small DM masses) or have $v^*$ close to, but smaller than, the maximum DM velocity in the lab frame $v_{\rm max}$ (for large DM masses). Intriguingly, this suggests that daily modulation, like annual modulation, is driven by the high-velocity tail of the DM velocity distribution. The centrality of these kinematic relations has recently been noted by \cite{Hochberg:2021pkt}, which discusses $v^*$ in the context of maximizing the total rate, and by \cite{Coskuner:2021qxo}, which uses a related $v^*$ to study modulation in the context of single-phonon production.

This paper is organized as follows. In Sec.~\ref{sec:Model}, we review the quantum chemistry relevant for describing the molecular orbitals in t-stilbene. In Sec.~\ref{sec:crystal} we describe the crystal structure of t-stilbene and justify our use of the isolated-molecule orbitals based on experimental measurements of the absorption and emission spectra. In Sec.~\ref{sec:Rate} we set up our calculation of the DM scattering rate, including the relevant molecular form factors and the daily modulation of the velocity distribution. In Sec.~\ref{sec:Results} we determine the daily modulation signal from DM in the Standard Halo Model (SHM) and derive a convenient test statistic for daily modulation in the presence of a non-modulating background. In Sec.~\ref{sec:Modulation} we study the kinematic features of t-stilbene which lead to a large modulation amplitude, and explore how a system with a different $v^*$ could lead to large daily modulation even for DM up to the GeV scale. We conclude in Sec.~\ref{sec:Conclusions}.

\section{Molecular Orbital Model}
\label{sec:Model}

\begin{table*}
\centering
\begin{minipage}{\textwidth}
\begin{tabular}{c | c c c c l l l l l}
$s$	& Platt Symbol &  Symmetry	& $ \Delta E \, [\!\eV]$ & Configuration amplitudes   \\ \hline 
$s_1$ & $^1B$  & $B_u$ 		& 4.240 	& $d_{7,8} = 0.94$, & $d_{4,11} = -0.24$ && &&	\Tstrut\\
$s_2$ & $^1G^-$ & $B_u$ 	& 4.788 	& $d_{7,10} = 0.53$,& $d_{5,8} = 0.53$, &&$d_{6,11} = 0.37$, &&$d_{4,9} = -0.37$	 \\
$s_3$ & $^1G^-$ & $A_g$ 	& 4.800 	& $d_{7,9} = 0.53$, & $d_{6,8} = 0.53$, &&$d_{5,11} = 0.37$, &&$d_{4,10} = -0.37$	\\
$s_4$ & $^1(C,H)^+$ & $A_g$ 	& 5.137 	& $d_{7,11} = 0.41$, &$d_{5,9} = -0.41$, &&$d_{6,10} = -0.41$, && $d_{4,8}= -0.59$	\\
$s_5$ & $^1H^+$ & $B_u$ 	& 5.791 	& $d_{5,10} = 0.54$, & $d_{6,9} = 0.54$, &&$d_{7,12} = 0.33$, &&$d_{3,8} = 0.33$	\\
$s_6$ & $^1G^+$ & $A_g$ 	& 6.264 	& $d_{7,9} = 0.68$, & $d_{6,8} = -0.68$ && && 	\\
$s_7$ & $^1C^-$ & $A_g$ 	& 6.013 	& $d_{7,11} = 0.66$, & $d_{4,8} = 0.54$, && &&	\\
$s_8$ & $^1G^+$ & $B_u$ 	& 6.439 	& $d_{7,10} = 0.65$, & $d_{5,8} = -0.65$ && &&	\\
\end{tabular}
\caption{The first eight excited states $s_{n = 1 \ldots 8}$, with their energy eigenvalues $ \Delta E(s_n)$ with respect to the ground state and coefficients $d^{(n)}_{ij}$ as calculated by Ting and McClure~\cite{ting1971}. \label{table:ting}}
\end{minipage}
\end{table*}

\begin{figure}
    \centering
    \includegraphics[width=65mm]{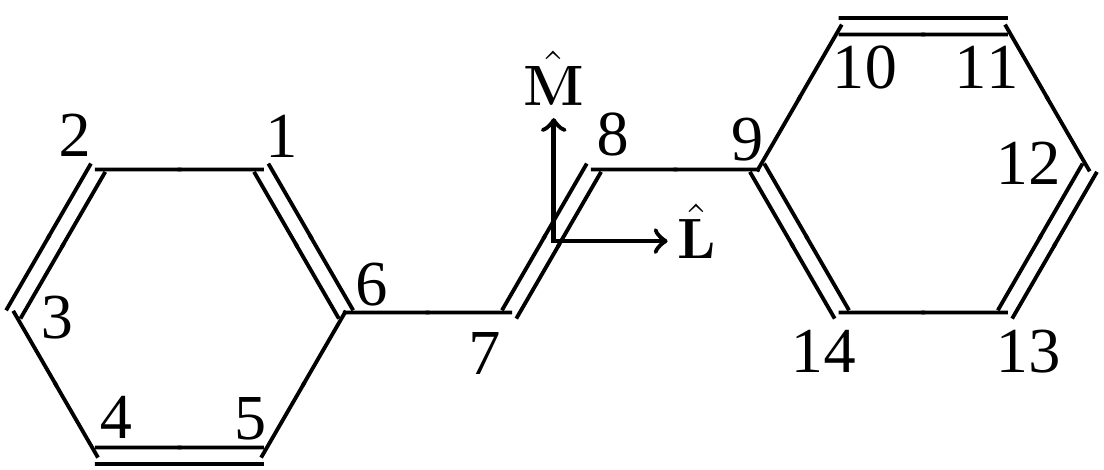}
    \includegraphics[width=65mm]{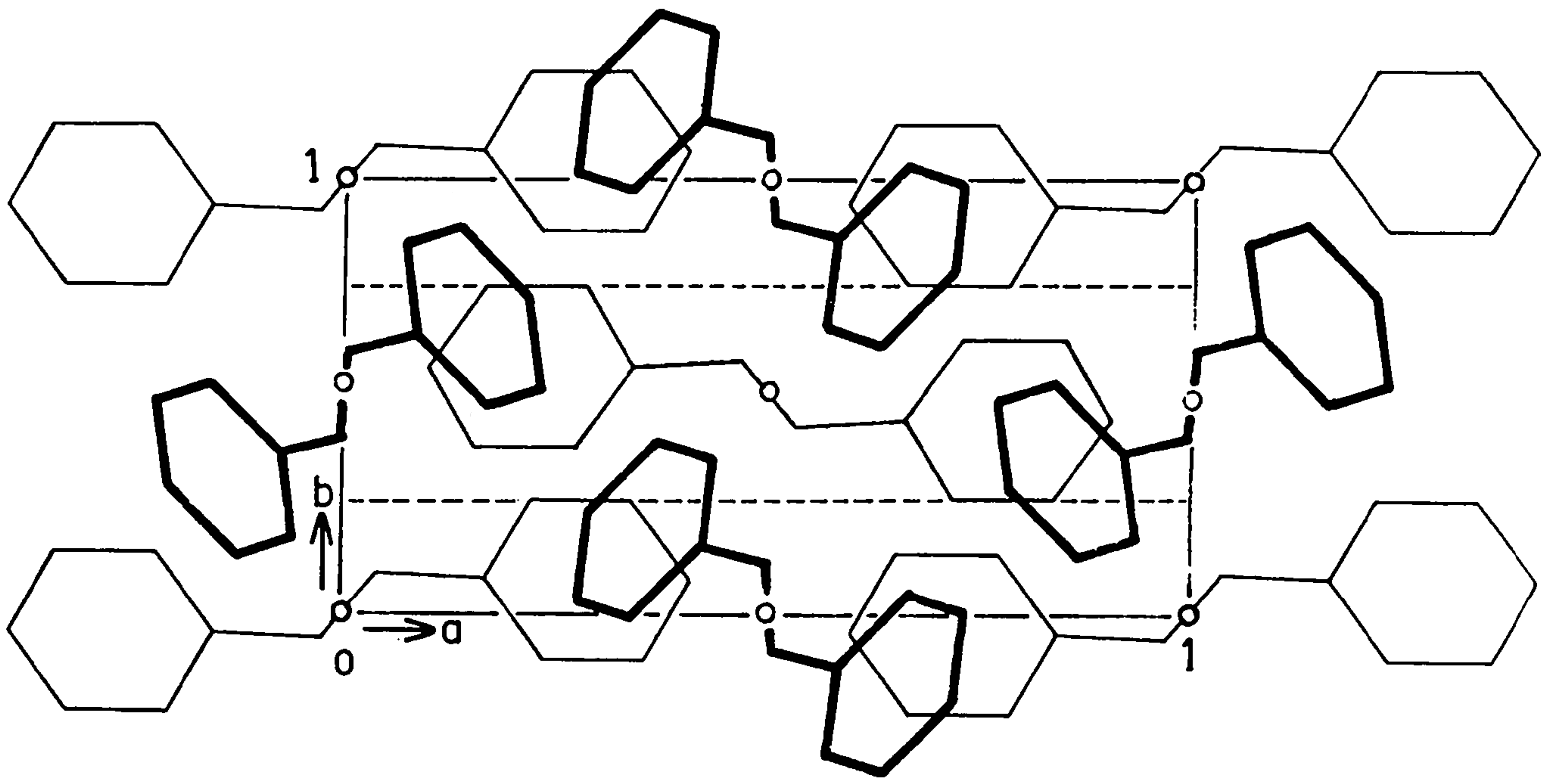}
    \caption{\textbf{Top:} The chemical  structure of trans-stilbene. Following the convention common in organic chemistry, vertices are taken to be carbon atoms, single lines are carbon--carbon single bonds, and double lines are carbon--carbon double bonds. Our numbering convention for the atoms is shown at each vertex, along with the $\hat{{L}}$ and $\hat{{M}}$ unit vectors used in our coordinate system.
    \textbf{Bottom:} A diagram of the unit cell of the trans-stilbene crystal, adopted from Ref.~\cite{Hoekstra1975}. Here $b$ is the axis of symmetry for the crystal: the positions of the molecules in the middle row are related to those of the upper or lower rows  by a translation of $\frac{1}{2} b$ and a rotation of $180^\circ$ about the $b$ axis. The long axis of each molecule ($\hat{{L}}_i$) is not perfectly perpendicular to the $b$ crystal axis. Molecules $(1)$ and $(-1)$ are shown with thin lines, $(2)$ and $(-2)$ with thick lines.
    \label{fig:struct} }
\end{figure}

Here we determine the many-electron wavefunctions for the electronic transitions which are responsible for the electronic-optical properties of trans-stilbene, following methods appropriate for all aromatic molecules. As shown in Fig.~\ref{fig:struct} (top), t-stilbene is an alternant hydrocarbon hosting two phenyl groups joined by an ethene bridge in the trans configuration (trans-1,2-diphenylethylene). This 14-carbon molecule is planar in the solid state and presents $C_{2h}$ molecular symmetry~\cite{Beveridge1965,Molina1997}, which is a $Z_{2}\times Z_{2}$ symmetry group composed of a two-fold symmetry axis, a center of inversion, a horizontal mirror plane, and the identity. Since this is the symmetry of the Hamiltonian, the electronic states of t-stilbene should transform as irreducible representations of $C_{2h}$. In order to construct electronic states which accurately describe the energy eigenstates of t-stilbene, we first construct the H$\text{\"u}$ckel molecular orbitals (HMOs) using a simple Linear Combination of Atomic Orbitals (LCAO) model taking into account only direct bonding interactions. We then take into account the configurational interactions and construct fully antisymmetric many-body states following the method of Pople, Pariser, and Parr (PPP)~\cite{pople1955electronic,Pariser1953,Pariser1953a}.

The HMOs, $\Psi_i$, of t-stilbene are constructed as linear combinations of Slater-type atomic orbitals (SAOs)
\begin{equation}
    \Psi_i = \sum_{j=1}^{14} c_{i}^{j} \phi_{2p_z}(\mathbf{r}-\mathbf{R}_j),
    \label{eq:LCAO}
\end{equation}
where $c_i^j$ are the coefficients to be determined, $\phi_{2p_z}$ are the atomic orbitals, and $\mathbf{R}_i$ are the equilibrium locations of the carbon nuclei using the numbering conventions in Fig.~\ref{fig:struct}. The $2p_z$ Slater atomic orbital is parameterized as 
\begin{equation}
\label{eq:phidef}
    \phi_{2 p_{z}}(\mathbf{r})=\sqrt{\frac{Z^{5}_\text{eff}}{2^5   \pi a_{0}^{3}}}  \frac{ r \cos \theta}{a_{0}} \exp \left(-\frac{Z_{\text {eff }} r}{2 a_{0}}\right),
\end{equation}
where $a_0=(\alpha m_e)^{-1}$ is the Bohr radius and $Z_\text{eff} = 3.15$ is the effective nuclear charge of the carbon $2p_z$ orbital \cite{Pariser1953}. The HMOs diagonalize the $14 \times 14$ core Hamiltonian matrix, $H_{l m} = \langle \phi_l |\mathcal{H}_{core}| \phi_m \rangle$, where $\langle \vec{r} | \phi_m \rangle = \linebreak \phi_{2p_z} (\vec{r} - \vec{R}_m)$ and only bonding atoms interact. Diagonalization is done by solving the following system,
\begin{equation}
    \sum_{j=1}^{14}\left[\left(H_{l j}-E_m \delta_{l j}\right) c_{m}^{j}\right]=0, \text { for } m=1,2, \ldots, 14.
\end{equation}
The core Hamiltonian contains two types of matrix elements; diagonal on-site energies and off-diagonal interaction energies, with values given in Appendix~\ref{sec:OrbitalDetails}. The onsite energy is an empirical quantity which is determined by the atomic species, while the off-diagonal energy is a measure of the bonded nuclear interactions determined by the effective charge and bond length. 
% The values of these energies are given in Appendix~\ref{sec:OrbitalDetails}. 
The ambiguity in the construction of degenerate states is resolved by requiring that all electronic states transform as irreducible representations of the $C_{2h}$ point group. Since the HMOs are constructed from 14 SAOs, diagonalization of the core Hamiltonian results in 14 HMOs $\Psi_1, \dots, \Psi_{14}$, numbered in order of increasing energy (up to degeneracies). See \cite{Blanco:2019lrf} for an example of this procedure performed on the simpler benzene molecule containing only six carbon atoms.

\begin{table*}
\centering
\begin{minipage}{\textwidth}
\begin{tabular}{c|rl c rl} 
Molecule    & & $(L_x, L_y, L_z)$ &&& $(M_x, M_y, M_z)$ \\ \hline
    $(1)$ & $\hat{L}=$  & $ ( 0.153, 0.988, - 0.022 )$ & {\color{white}.}\hspace{1cm}{\color{white}.}& $\hat{M}=$ & $  (0.467, - 0.068, 0.882 )$ \Tstrut\\
    $(2)$ & $\hat{L}=$ & $ ( -0.809, 0.565, 0.162)$ && $\hat{M}=$ & $  (- 0.130, - 0.458, 0.879 )$ \Tstrut\\
    $(-1)$ & $\hat{L}=$ & $(  -0.153, - 0.988, - 0.022 ) $&& $\hat{M}=$ & $  (-0.467 , 0.068, 0.882 )$ \Tstrut\\
    $(-2)$ & $\hat{L}=$ & $ (0.809, - 0.565 , 0.162 )$ && $\hat{M}=$ & $  (0.130 , 0.458, 0.879 )$ \Tstrut\\
         \hline
\end{tabular}
    \caption{The vectors $\hat{L}$ and $\hat{M}$ describe the orientations of each of the four molecular constituents of the unit cell, shown here in the right-handed crystal basis $\hat{x}= \hat{c}$, $\hat{y}= \hat{a}'$, $\hat{z}= \hat{b}$.}
    \label{table:geometry}
\end{minipage}
\end{table*}
 
In order to form antisymmetric many-electron wave functions, we take Slater determinants of the filled HMOs. The ground state $| g \rangle$ is approximately given by the following combination of orbitals,
\begin{equation}
   \psi_G = |\Psi_{1} \overline{\Psi}_{1}\Psi_{2} \overline{\Psi}_{2}\Psi_{3} \overline{\Psi}_{3}\Psi_{4} \overline{\Psi}_{4}\Psi_{5} \overline{\Psi}_{5}\Psi_{6} \overline{\Psi}_{6}\Psi_{7} \overline{\Psi}_{7}|,
\end{equation}
where $|\cdots|$ denotes the antisymmetrized product of the HMOs and $\overline{\Psi}_i$ is the opposite spin state as $\Psi_i$. Following standard conventions in quantum chemistry, we label some $Z_2$ symmetries of the many-electron wavefunction by $A(B)$ and $g(u)$, corresponding to (anti)symmetry with respect to transformation under $180^\circ$ rotation about the $z$-axis normal to the molecular plane and inversion through the center of mass, respectively. Notice that the ground state represents an electronic configuration in which the lowest 7 HMO's are filled by pairs of electrons in the spin-singlet configuration. Therefore, the ground state transforms as the $A_{g}$ representation, being totally symmetric under the transformations in the $C_{2h}$ group. This is a generic feature of the ground state of alternant hydrocarbons. Reduction of the 14-dimensional t-stilbene representation of $C_{2h}$ predicts 7 $A_{g}$ and 7 $B_{u}$ states~\cite{Beveridge1965}.  These states correspond to the multi-electron configurations of the HMOs which each have either $A_{u}$ or $B_{g}$ symmetry~\cite{Molina1997}.

We construct the  multi-electron states starting with the $A_{g}$ ground state and proceeding upwards in energy through the one-electron singlet excitation configurations, $\psi_i^j$ as follows,
\begin{equation}
    \psi_i ^j = \frac{1}{\sqrt{2}} (|\Psi_{1} \overline{\Psi}_{1}...\Psi_{i} \overline{\Psi}_{j}...\Psi_{7} \overline{\Psi}_{7}| - |\Psi_{1} \overline{\Psi}_{1}...\Psi_{j} \overline{\Psi}_{i}...\Psi_{7} \overline{\Psi}_{7}|).
\end{equation}
The electronic repulsion is taken into account by appending the electronic two-body interaction to the core H$\text{\"u}$ckel Hamiltonian:
\begin{equation}
    \mathcal{H}_{\rm PPP} = \mathcal{H}_{core} + \sum_{\langle ij \rangle}{\frac{4\pi \alpha}{r_{ij}}},
\end{equation}
where $\alpha$ is the fine-structure constant, $r_{ij} = |\vec{r}_i - \vec{r}_j|$, and the sum runs over all pairs of electrons $\langle ij \rangle$ with $i \neq j$. This PPP Hamiltonian perturbs the energy levels of the $\psi_i^j$ configurations and mixes degenerate states of like symmetries to produce PPP HMOs of either $A_{g}$ or $B_{u}$ symmetry~\cite{Beveridge1965,Molina1997,ting1971}. The PPP energy eigenstates are expressed as a linear combination of $\psi_i^j$,
\begin{align}
| s_n \rangle = \sum_{i,j>i} d^{(n)}_{ij} | \psi_{i}^j \rangle ,
&&
\sum_{ij} | d^{(n)}_{ij} |^2 = 1.
\end{align}
The leading coefficients $d^{(n)}_{ij}$ (known as configuration amplitudes) for the first $n=1 \dots 8$ excited states of t-stilbene are tabulated in Tab.~\ref{table:ting}~\cite{ting1971}. To better match our analysis to experimental data, we take the experimentally-determined energy eigenvalues, which are also listed in Tab.~\ref{table:ting}~\cite{ting1971}.

The spin-singlet configurations we have focused on are responsible for the radiative de-excitations known as fluorescence that could be seen with single-photon detectors in a DM experiment. The triplet states are classically forbidden from decaying down to the ground state and hence are responsible for the delayed fluorescence component of photoluminescence which generically has a significantly lower quantum yield~\cite{1964a}.

\section{Crystal Structure}
\label{sec:crystal}
In the solid state, t-stilbene is a monoclinic crystal belonging to the space group $C5_{2h}(P2_{1}/c)$, with unit cell parameters $a=12.29$ \AA, $b=5.66$ \AA, $c=15.48$ \AA, and $\gamma=112^{\circ}$~\cite{Hoekstra1975}. In this convention, the crystal coordinate basis is defined by a unit vector $\hat{b} = \vec{b}/b$ that is orthogonal to both $\hat{a}$ and $\hat{c}$, and $\gamma$ is the opening angle between $\hat{a}$ and $\hat{c}$. The coordinate system is right-handed, such that $\vec{c}\times \vec{a} \parallel \vec{b}$. To form an orthonormal basis we define an $\hat{a}'$ unit vector, $\hat{a}' = \hat{b} \times \hat{c}$. The crystal is symmetric with respect to translations of $\vec{a}$ and $\vec{c}$, and to the twofold screw action composed of the translation $\vec{b}/2$ with $180^\circ$ rotation about $\hat{b}$.   

Four distinct molecules of t-stilbene inhabit each unit cell of the crystal (see Fig.~\ref{fig:struct}, bottom): an $M_1$ and $M_2$ with different orientations, and an $M_{-1}$ and $M_{-2}$, which are the images of molecules $M_{1,2}$ (respectively) under the twofold screw action along $\vec{b}$~\cite{Robertson1937}. In Tab.~\ref{table:geometry} we provide the orientations of each of the four molecules in terms of their unit vectors $\hat{L}$ and $\hat{M}$ identified in Fig.~\ref{fig:struct}, in a crystal coordinate system where the $\hat{z}$ direction is assigned to the $\vec{b}$ symmetry axis. The position of each molecule within the crystal is listed in Refs.~\cite{Hoekstra1975,Robertson1937}, but the DM--stilbene scattering rate depends only on their rotational orientation because the kinematics of the scattering process do not permit coherent scattering over an entire unit cell (see Secs.~\ref{sec:Rate} and~\ref{sec:Results} below). 

The molecular orbital model derived in the preceding section is empirically valid for macroscopic single-crystal samples of t-stilbene. Although lattice effects are known to perturb the energy bands of molecular crystals~\cite{davydov1953theory}, the Davydov splitting of the molecular bands, which is due to the dipole-dipole (and to a lesser extent quadrupole-quadrupole) interaction of neighboring molecules in a molecular crystal, is known to be very small in single-crystal t-stilbene~\cite{Chaudhuri1969}. Furthermore, the lowest two UV-absorption bands (labeled A and B in the literature) of t-stilbene in the liquid state remain relatively untouched when observed in the solid state~\cite{Suzuki_1960}. The A band is thought to arise from the $g \to s_1$ through $g \to s_4$ transitions while the B band is thought to arise from the $g \to s_5$ transition~\cite{Beveridge1965}. These bands are a direct measurement of transitions between the many-electron configurations described in the previous section which represent the non-interacting single-molecule electronic state, usually taken to be most like the molecular environment of the low-temperature liquid  or gas state. Since these bands remain the same in the solid states, we conclude that the PPP model accurately describes the molecules of t-stilbene in the solid state where lattice effects are only very weak.

A mole of t-stilbene has a mass $180.24$ g and occupies a volume of $185.69 \cm^{3}$. Thus, a kilogram of detector material can be fabricated from a cube of t-stilbene of 10.1 cm per side, or a 1 cm thin sheet of approximately one square foot. These sizes will be convenient to instrument with conventional photodetectors, such as photomultiplier tubes or charge-coupled devices (CCDs).

\section{Rate Calculation}
\label{sec:Rate}

The interaction of DM with the electronic system of a molecule produces a detectable signal through a process analogous to photoluminescence. This process can be separated into two stages: electronic excitation from DM-electron scattering followed by radiative deexcitation. Specifically, DM-electron scattering will produce detectable scintillation photons from t-stilbene at a rate given by \cite{Blanco:2019lrf}
\begin{widetext}
\begin{equation}
    R = \frac{\Phi_{FB} N_\text{A} m_T}{m_{\rm mol}^\text{(t-stil)}} \frac{\rho_\chi}{m_\chi} \frac{\bar\sigma_e}{\mu_{\chi e}^2} \sum_{i = 1} \int\! \frac{d^3 \vec{q}}{4\pi} \int\! d^3 \vec{u} \, f_\chi(\vec{u})  \delta\left( \Delta E(s_i) + \frac{q^2}{2 m_\chi} - \vec{q} \cdot \vec{u} \right)F_\text{DM}^2(q) |f_{g \rightarrow s_i}(\vec{q})|^2.
\label{eq:ratedef}
\end{equation}
\end{widetext}
Here, $N_{\rm A} \simeq 6.022 \times 10^{23}$ is Avogadro's number; $m_T$ is the total detector mass; the molar mass of t-stilbene is $m_{\rm mol}^\text{(t-stil)}= 180.25$ g; $\rho_\chi$ and $m_\chi$ are the DM mass density and mass, respectively; $\mu_{\chi e}$ is the DM--electron reduced mass; $\bar{\sigma}_e \equiv \frac{\mu_{\chi e}^2}{16 \pi m_\chi^2 m_e^2}\langle |\mathcal{M}(q_0)|^2 \rangle$ is a fiducial DM--electron cross section proportional to the free-particle spin-averaged squared matrix element $\langle |\mathcal{M}|^2 \rangle$ for $\chi-e$ scattering, evaluated at $q_0 = \alpha m_e$; $f_\chi(\vec{u})$ is the DM velocity distribution in the lab frame; $F_{\rm DM}(q)$ is a form factor parameterizing the fundamental DM-electron interactions, which has the limits $F_{\rm DM}(q) = 1$ for a contact interaction and $F_{\rm DM}(q) =(\alpha m_e/q)^2$ for a long-range interaction; and $\Delta E(s_i)$ is the excitation energy for each $s_i$ above the ground state as given in Tab.~\ref{table:ting}. 

The detector-dependent quantities are the molecular form factor $|f_{g \rightarrow s_i}(\vec{q})|^2$, representing the transition amplitude from the ground state to a singlet excited state $s_i$, and the bulk fluorescence quantum efficiency $\Phi_{FB}$, representing the probability that a molecular excitation will produce a photon through radiative deexcitation that will exit the detector without being absorbed. As is the case with photoluminescence, the emission lines are broadened by vibrational energy sublevels, thermal motion, and lattice effects. The emission spectrum of the single crystal is a continuum of peaks which closely resembles the molecular and micro-crystalline emission spectra but is modified by self-absorption and lattice effects~\cite{Birks1954}. Here we focus on computing the form factor and the quantum efficiency, which determine the signal rate, leaving a detailed investigation of the emission spectrum (which determines the signal photon wavelength and hence the detection mechanism) for future work. We compute the molecular form factors using the first 8 singlet transitions since these are the transitions responsible for the first three lowest-lying absorption bands of trans-stilbene~\cite{Beveridge1965}. Since the probability of interaction is suppressed for higher energy thresholds (see Fig.~\ref{fig:rateByTransition}), it is expected that the rate will be driven primarily by the strongest low-lying excitations.  Using a simple, spherically-symmetric DM velocity distribution, the SHM, we also calculate daily modulation effects due to the rotation of the Earth. 

\subsection{Molecular Form Factors}

\begin{figure*}
    \centering
    \includegraphics[width=0.98\textwidth]{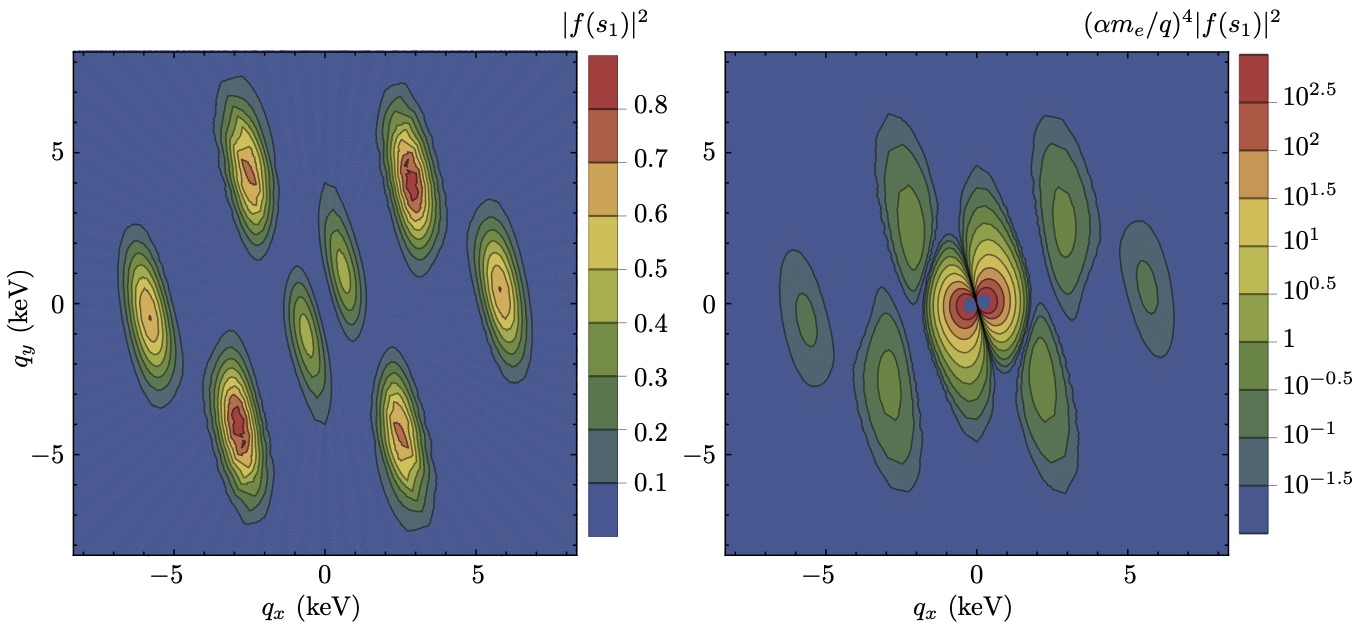}
    \caption{Molecular form factors for the $g\to s_1$ transition for momentum transfers $\vec{q} = (q_x, q_y, 0)$ in the plane of the molecule. \textbf{Left:} contour plot of $|f_{g \rightarrow s_1}|^2$, showing the hexagonal symmetry of the benzene ring with secondary inner peaks arising from the extended structure of the joined benzene rings. \textbf{Right:} the product of $|f_{g \rightarrow s_1}|^2$ with $F_\text{DM}^2$ for $F_\text{DM} \propto 1/q^2$, proportional to the rate integrand \eqref{eq:RTotg0} for a light mediator (logarithmic color scale). }
    \label{fig:formFactors}
\end{figure*}

The molecular form factor as calculated for PPP configurational states $s_n$ is given by the following,
\begin{align}
f_{g \to s_n}(\vec{q}) &=  \left\langle \psi_{s_n} (\vec{r}_1 \ldots \vec{r}_{14}) \left|  \sum_{m=1}^{14} e^{i \vec{q} \cdot \vec{r}_m } \right| \psi_G(\vec{r}_1 \ldots \vec{r}_{14}) \right\rangle \nonumber \\
&=\sum_{ij} d^{(n)}_{ij} 
 \left\langle \psi_i^j \left|   e^{i \vec{q} \cdot \vec{r} } \right| \psi_G \right\rangle  \nonumber \\
&=\sqrt{2} \sum_{ij}  d^{(n)}_{ij} \langle \Psi_j(\vec{r}) |e^{i\vec{q} \cdot \vec{r}}|\Psi_i (\vec{r})\rangle.
 \label{eq:fgs}
\end{align}
where $\vec{r}_m$ is the position of electron $m$. In the second line we have isolated the contribution from the singlet states which only contain a single-electron excitation above the ground state, and in the third line we have transformed to the basis of HMOs, where the factor of $\sqrt{2}$ is effectively a spin degeneracy factor. The matrix element of HMOs may be readily computed from Eq.~(\ref{eq:LCAO}) in momentum space, where the wavefunctions are simply the momentum-space $2p_z$ orbitals times a product of phase factors determined by the positions of the carbon atoms \cite{Blanco:2019lrf}. An example of the single-molecule form factor for the $g \to s_1$ transition is shown in Fig.~\ref{fig:formFactors} (see also App.~\ref{sec:OrbitalDetails}), showing strong damping of the form factor beyond a characteristic momentum scale given by $q^* \simeq Z_{\rm eff}/(2 a_0) \simeq 6 \ {\rm keV}$. There are also secondary ``inner'' peaks at $q \simeq 2\pi/\ell \simeq 1.2 \ {\rm keV}$ where $\ell \approx 0.83 \; \text{nm}$ is the length scale of the long axis of t-stilbene.

Measurements of the spectrum of trans-stilbene observe three absorption bands, $A$, $B$ and $C$, which have been identified primarily with the $s_1$, $s_5$ and $s_8$ molecular transitions, respectively~\cite{Beveridge1965}. 
% This motivates our choice to consider only the first eight excited states in our analysis. 
Compared with the $B$ and $C$ bands, the $A$ band is larger in magnitude and broader in frequency space, overlapping with the $s_{2 \ldots 4}$ transitions. Our analysis for dark matter scattering reproduces these features: for example, the $s_1$ transition dominates the scattering rate, comprising 50--70$\%$ of the rate both near-threshold and at large $m_\chi$. If $m_\chi$ is large enough that the higher-energy excitations are kinematically accessible, the scattering rate receives secondary contributions from $s_3$ and $s_4$, and typically smaller contributions from $s_2$, $s_5$ and $s_8$. The $s_6$ and $s_7$ transition rates remain negligibly small at all values of $m_\chi$. We provide additional detail regarding the separate molecular transitions that contribute to the scattering rate in Appendix~\ref{sec:ffdetails}.

In a t-stilbene crystal, the unit cell contains four molecules of different orientations, as described in Section~\ref{sec:crystal}. As discussed in more detail below, the momentum transfers required to deposit energy above $\Delta E(s_1)$ are sufficient to localize the interaction to a single molecule within a unit cell, so to compute the rate we treat the scattering as incoherent between different molecules, and we sum over four different squared form factors $|f^{(i)}_{g \to s}(\vec{q})|^2$ rotated to give the appropriate orientations of each molecule with respect to $\vec{q}$. The interaction rate of DM with a crystal of t-stilbene then scales like the product of the rate calculated as described with the total number of unit cells $N_{\rm uc} = N_{\rm mol}/4$ in the entire crystal, which is proportional to the total crystal mass $m_T$.

\subsection{Quantum Efficiency}
 Given the molecular fluorescence quantum efficiency ($\Phi_F$), defined as the ratio of emitted photons to absorbed, the probability $\Phi_{FB}$ of a photon exiting the bulk target after an excitation is then given by
 \begin{equation}
   \Phi_{FB} = (1-a_{xx})\Phi_F,
 \end{equation}
where $a_{xx}$ is the probability of self absorption. At liquid nitrogen temperatures, $\Phi_F \simeq 97\%$~\cite{Charlton1977,Birch1976,Sharafy1971}, approaching unity at cryogenic temperatures. Furthermore, using the photoluminescent spectra of t-stilbene in the liquid, microcrystalline, and macroscopic single-crystal state as measured via reflection and transmission, Birks et al.~conclude that $a_{xx} \approx 0.35$ as the continuous reabsorption and emission of the photon gradually red-shifts the radiation into wavelengths to which the bulk crystal is transparent~\cite{Birks1954}. Thus, the bulk fluorescence quantum efficiency of t-stilbene is at least $\Phi_{FB} = 0.63$~\cite{Katoh2009,Zaitseva2015}, likely approaching $0.65$ at cryogenic temperatures, though we use the lower value to be conservative. We propose such a detector to be run around 100K in order to maximize bulk quantum efficiency while maintaining a high enough temperature to run CCD based photo-detectors~\cite{Tiffenberg:2017aac}.

\subsection{DM Velocity Distribution}
\label{sec:Velocity}
We denote the velocity of a particle in the Milky Way reference frame by $\vec v$. We adopt the SHM ansatz for the bulk of the Milky Way DM distribution. The DM velocity is then distributed according to $f_0(\vec v) = \exp(- |\vec v|^2/2\sigma_0^2) \Theta(|\vec v|^2 - v_{\rm esc}^2)/N_0,$ where the dispersion $\sigma_0$ is related to the velocity of the Local Standard of Rest by $\sigma_0 = v_0/\sqrt2$ and the normalization is
\begin{equation} \label{vel-N0}
    N_0 \! = \! \pi^{3/2} v_0^3 \left[ \erf \left( \! \frac{v_{\rm esc}}{v_0}\! \right) - \frac2{\sqrt\pi}\frac{v_{\rm esc}}{v_0} \exp \! \left( \! -\frac{v_{\rm esc}^2}{v_0^2} \! \right)\! \right].
\end{equation}
Numerically, the Local Standard of Rest has value $v_0 \simeq 220$ km/s and the escape velocity is near $v_{\rm esc} \simeq 544$ km/s \cite{Evans:2018bqy}; the uncertainties on these values are nonzero, but $\lesssim 10\%$. We use this velocity distribution in order to facilitate comparison with previous studies, but future exploration of the implications of more realistic velocity distributions will be important for interpreting any future experimental results.

\begin{figure*}
    \centering
    \includegraphics[width=0.98\textwidth]{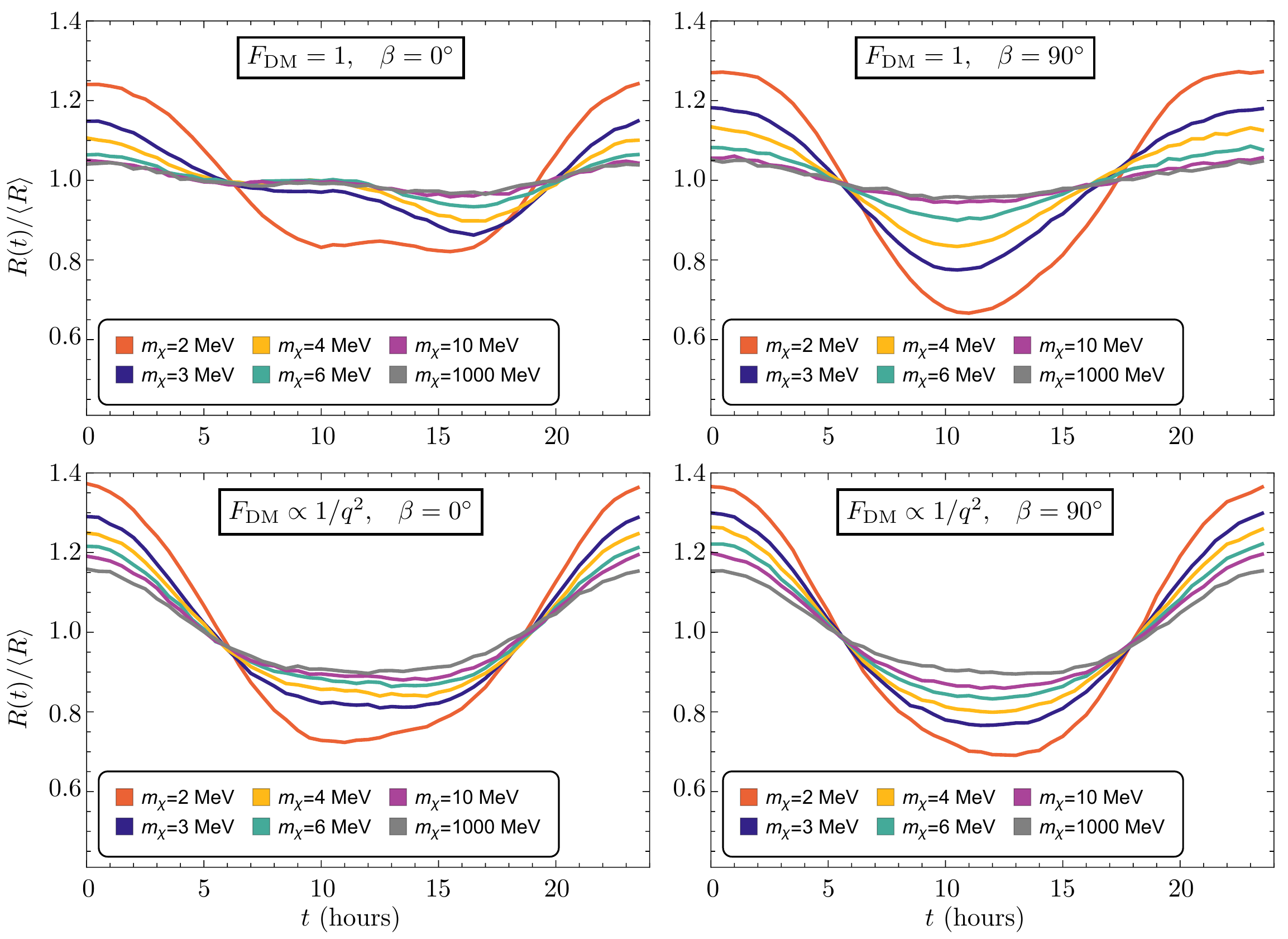}
    \caption{Normalized modulation signals for a variety of DM masses, $m_\chi = \text{2--1000}\,\MeV$, for a crystal in $\beta=0^\circ$ and $\beta=90^\circ$ orientations. Above $10\,\MeV$, the rate relaxes into a function of time that is nearly independent of the DM mass and with modulation amplitude only mildly dependent on the crystal orientation. The peak-to-trough modulation amplitudes are as large as 60\% at low masses and 10\% at high masses for $F_\text{DM}=1$, increasing to  70\% at low masses and 25\% at high masses for $F_\text{DM} = (\alpha m_e/q)^2$.
    }
    \label{fig:rateNorm}
\end{figure*}

To calculate the rate in \eqref{eq:ratedef}, we integrate over the velocity measured in the laboratory $\vec u$, which is related to the velocity in the Milky Way frame by $\vec v = \vec u + \vec v_\oplus$, where $\vec v_\oplus(t)$ is the velocity of the Earth as measured in the Milky Way frame. Following the conventions of \cite{Coskuner:2019odd}, we use the energy conservation $\delta$-function to resolve the integral over the velocity distribution:
\begin{align}
    g_0(\vec{q}, t) &\equiv \int d^3 \vec{u}\, f(\vec v_\oplus(t) + \vec u) \, \delta\left(\Delta E - \vec q \cdot \vec u + \frac{q^2}{2m_\chi} \right) \nonumber
    \\ & = \frac{\pi v_0^2}{q N_0} \left( e^{-v_-^2/v_0^2} - e^{-v_{\rm esc}^2/v_0^2} \right),  \label{vel-g}
\end{align}
where
\begin{equation}
    v_-(\vec{q}, t) = \min \! \left( \! v_{\rm esc}, \frac{\Delta E}{q} + \frac{q}{2m_\chi} + \vec v_\oplus(t) \cdot \vec{\hat q} \right).
    \label{eq:vMinus}
\end{equation}
The momentum transfer $\vec q = q \vec{\hat q}$ has been related to the velocity by the requirement that $\Delta E = \vec q \cdot \vec u - q^2/(2m_\chi)$, and all dependence on the local velocity $\vec{v}_\oplus$ is contained in $v_-$.

The rotation of the Earth over a 24-hour period enters the rate by casting the direction of Earth's velocity (more precisely, the velocity vector of a fixed laboratory location on the Earth's surface) as a function of time \cite{Coskuner:2019odd}:
\begin{equation}
\vec{\hat v}_\oplus (t) \! = \! \!
\left( \begin{array}{c c c} \cos\beta & -\sin\beta & 0 \\  \sin\beta & \cos \beta & 0 \\ 0 & 0 & 1 \end{array} \right) \!
\left( \! \begin{array}{c} \sin\theta_e \sin \vartheta \\ \sin\theta_e \cos\theta_e ( \cos \vartheta - 1) \\ \cos^2 \theta_e + \sin^2 \theta_e \cos\vartheta \end{array} \! \right),
\label{eq:vE}
\end{equation}
where $\vartheta(t) = 2\pi \times \left( \tfrac{t}{24\, \text{h} } \right)$, 
$\theta_e \approx \, 42^\circ$, and we have chosen to align the $(x,y)$ plane of the crystal to be perpendicular to the direction of the DM wind at time $t=0$ with the initial orientation of the crystal with respect to rotations about the $\hat z$ axis given by $\beta$. As is clear from \eqref{eq:vMinus}, the only aspects of the Earth velocity vector that we need to know when calculating rates in the context of the SHM are the Earth's speed, for which we adopt $|\vec v_\oplus| \equiv v_\oplus = 234$ km/s, and the angle between the Earth's north pole and its velocity in the Milky Way frame, $\theta_e$. Our formalism is easily extended to other velocity distributions by making the substitution in \eqref{eq:vMinus} of $\vec{v}_\oplus \to \vec{v}_\oplus - \langle \vec v \rangle$ where $\langle \vec v \rangle$ is the mean velocity of the DM distribution as measured in the Milky Way frame. The vector $\langle \vec v \rangle$ is zero by definition for the SHM, but would be nonzero for substructure in the form of a stream.

In terms of $g_0$, the total (time-dependent) rate per unit mass is
\begin{equation}
    \frac{R(t)}{m_T} = \frac{\Phi_{FB} N_\text{A}}{m_{\rm mol}^\text{(t-stil)}} \frac{\rho_\chi}{m_\chi} \frac{\bar\sigma_e}{\mu_{\chi e}^2} \sum_{i = 1} \int\! \frac{d^3 \vec{q}}{4\pi} g_0(\vec{q}) F_\text{DM}^2(q) |f_{s_i}(\vec{q})|^2,
    \label{eq:RTotg0}
\end{equation}
where we emphasize that $g_0(\vec{q})$ is implicitly also a function of $\vec{v}_{\oplus}(t)$. 
Eqs.~(\ref{vel-g})--(\ref{eq:RTotg0}) indicate that, from the perspective of kinematics alone, the largest modulation will occur when $v_-(\vec{q})$ modulates around $v_{\rm esc}$. However, as we will see in Secs.~\ref{sec:Results} and~\ref{sec:Modulation} below, the morphology of the molecular form factors will also play a large role in determining the modulation.

\section{Daily modulation reach}
\label{sec:Results}

\begin{figure*}
    \centering
    \includegraphics[width=0.98\textwidth]{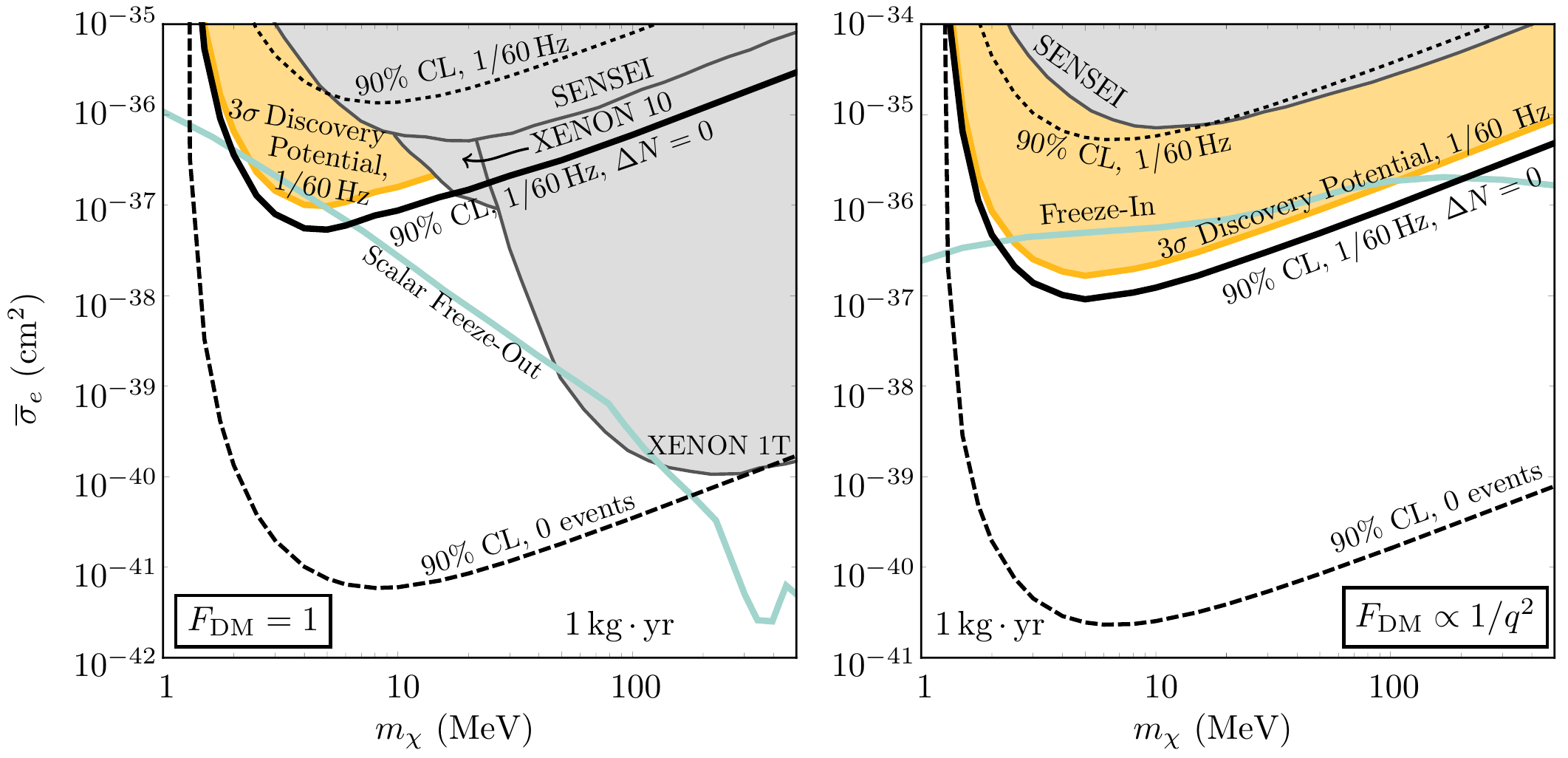}
    \caption{The capability of a $1\, \text{kg}\cdot \text{year}$ t-stilbene experiment to detect or exclude DM models with $F_\text{DM}=1$ (left) or $F_\text{DM} = (\alpha m_e/q)^2$ couplings to electrons, shown with existing limits from SENSEI~\cite{Barak:2020fql}, XENON~10~\cite{Essig:2017kqs}, and XENON~1T~\cite{Aprile:2019xxb}. The dotted and dashed lines show the 90\% CL exclusions that can be set from the total number of events, without considering modulation effects, for $R=1/60\,\text{Hz}\, \text{kg}^{-1}$ ($N_\text{events} \approx 5.26 \times 10^5$) and for $N_\text{events}=0$, respectively. The orange shaded regions indicate parameter space that leads to a sufficiently large modulation signal that a $1~\text{kg}\cdot\text{year}$ experiment could observe a $3 \sigma$ detection, given a total observed rate of $R=1/60\,\text{Hz}\, \text{kg}^{-1}$.  The solid black ``$\Delta N=0$'' lines show the improved limit that can be set from a null result exhibiting no daily modulation but the same total observed rate. Each plot also shows (in blue) a benchmark model from Ref.~\cite{Battaglieri:2017aum} as a target for the experimental sensitivity. In the $F_\text{DM}=1$ example the scalar DM abundance is set by freeze-out mediated by a dark photon of mass $m_{A'} = 3 m_\chi$,  while for $F_\text{DM}\propto 1/q^2$ we show freeze-in via light mediator, $m_{A'} \ll 3\,\keV$.
    }
    \label{fig:exclusionsSHM}
\end{figure*}

\begin{figure*}
\centering
    \includegraphics[width=0.98\textwidth]{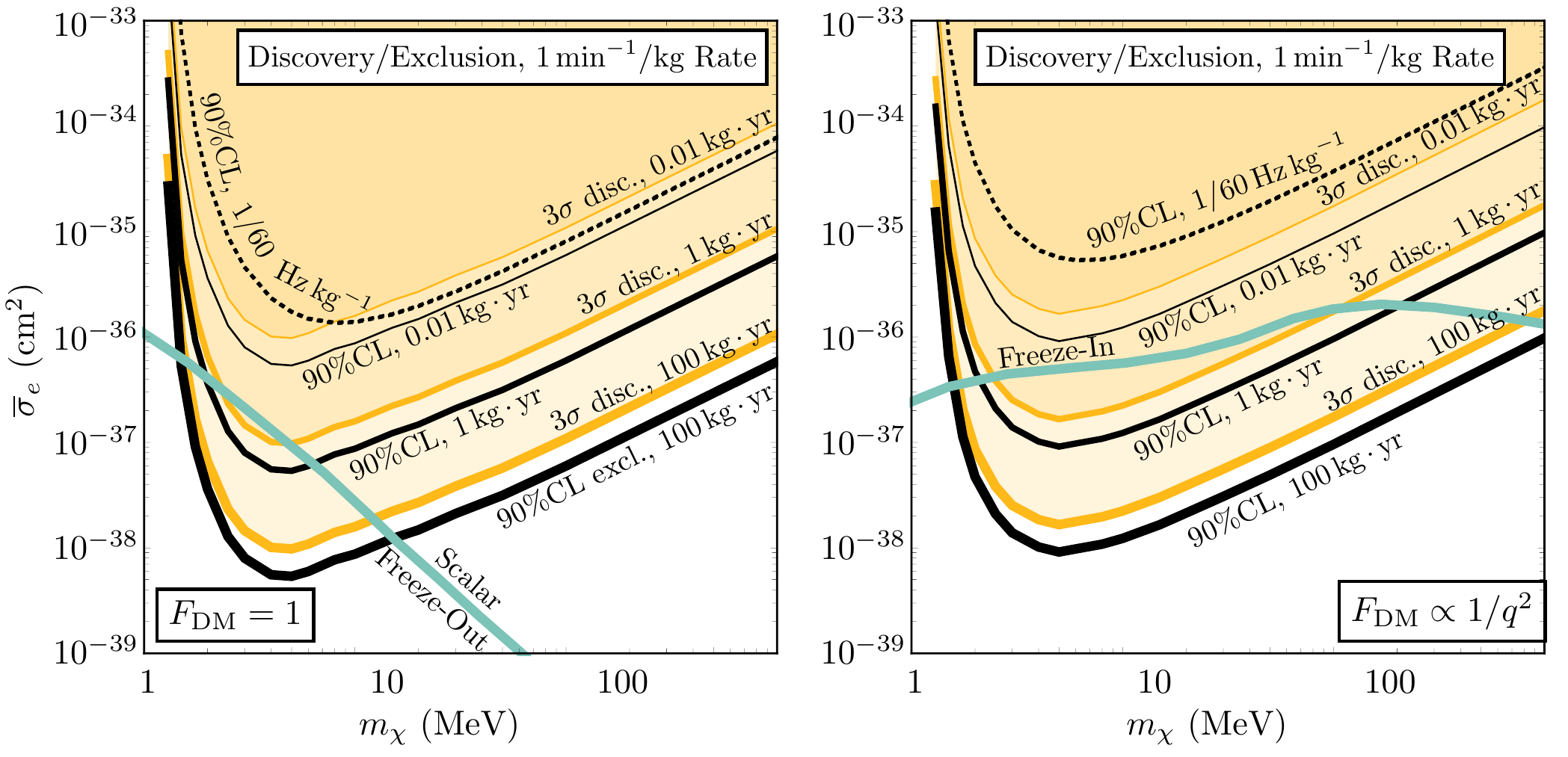}
    \caption{ As a demonstration of the utility of daily modulation, we show the $3\sigma$ discovery and 90\% CL exclusion potential for a trans-stilbene experiment with a background rate of $1/60\,\text{Hz}\,\text{kg}^{-1}$, for exposures of 0.01\,$\text{kg}\cdot\text{year}$, 1\,$\text{kg}\cdot\text{year}$, and 100\,$\text{kg}\cdot\text{year}$. The dashed lines, labelled ``90\% CL 1/60\,Hz\,$\text{kg}^{-1}$,'' show the 90\% CL exclusion from an analysis that does not consider the daily modulation effects. The inclusion of daily modulation in the statistical analysis allows even the $0.01\, \text{kg}\cdot\text{year}$ exposure to set a significantly stronger limit on $\bar\sigma_e$. For the $F_\text{DM}=1$ and $F_\text{DM} = (\alpha m_e/q)^2$ form factors we show the benchmark freeze-out and freeze-in models, respectively, from Ref.~\cite{Battaglieri:2017aum}.
    }
    \label{fig:discovery}
\end{figure*}

With the molecular form factors in hand, we can compute the total DM-induced excitation rate by summing over the eight lowest transitions $s_i$ for a given choice of velocity distribution and DM form factor. As we show in Appendix~\ref{sec:OrbitalDetails}, the lowest-energy transition $g \to s_1$ dominates both the daily modulation effect and the total average rate. Near the mass threshold, $m_\chi \lesssim 10 \, \MeV$, the $g \to s_3$ transition contributes at the $20\%$ level, with all other transitions contributing less than $10\%$ of the total rate. Above 10\,\MeV\ the $s_1$ transition remains dominant, accounting for about 50\% of the total rate for both DM form factors. With $F_\text{DM}=1$ and $m_\chi \gtrsim 100\,\MeV$ the $s_3$ and $s_4$ transitions contribute equally, at the 15\% level. For the same masses and $F_\text{DM}\propto 1/q^2$, the $s_3$ transition provides a larger 20\% correction to the total rate, compared to less than 15\% from $s_4$ and less than 10\% from each of the other excited states. This behavior is distinct from the case of benzene, where the lowest-energy transition is dipole-forbidden \cite{Blanco:2019lrf}.

Fig.~\ref{fig:rateNorm} shows the modulating rate $R(t)$ over a 24-hour period (one sidereal day) for two different alignment angles of the detector crystal, $\beta = 0^\circ$ and $\beta = 90^\circ$,
normalized by the average scattering rate,
\begin{align}
\langle R \rangle = (24\,\text{h})^{-1} \int_0^{24 \,\text{h} } \! dt\, R(t) .
\label{eq:rangle}
\end{align}
We see that the peak-to-trough modulation amplitude is as large as 60\% (10\%) for a low-mass (high-mass) DM particle interacting via a heavy mediator, climbing to 70\% (25\%) for a low-mass (high-mass) DM particle interacting via a light mediator. This is on the same scale, or larger than, the \emph{annual} modulation amplitude for WIMP-nuclear scattering well above threshold \cite{Drukier:1986tm, Freese:1987wu, Freese:2012xd, DelNobile:2015nua, Witte:2016ydc}, as well as for DM-electron scattering in semiconductors at high masses \cite{Lee:2015qva,Radick:2020qip}.

Assessing evidence in favor of a signal will be an important step in making a DM discovery, and the daily modulation is an important handle for improving our statistical power. As we discuss in more detail in Appendix \ref{sec:Stats}, the statistical significance that we formally assign to a modulating signal is
\begin{equation}
    \Delta L = - 2 \sum_k n_k \ln \left[\nu_k^m(\bm{\theta}^m)/\nu_k^0(\bm{\theta}^0) \right],
\label{eq:LMod}
\end{equation}
where $k$ labels the data bins, $\nu_k^m$ is the number of expected events in bin $k$ assuming a modulating signal, $\nu_k^0$ is the expected number of events in bin $k$ assuming a constant rate, and $\bm{\theta}^{m,0}$ are parameters describing the expected rate in the modulating and non-modulating scenario, respectively. The values of $\Delta L$ are distributed as a $\chi^2$ distribution of the number of additional degrees of freedom needed to characterize the modulating (as opposed to the non-modulating) signal; in the case of two bins, this would be a $\chi^2$ with two degrees of freedom. Although we focus on the two-bin case in the remainder of this analysis, we emphasize that \eqref{eq:LMod} is appropriate for any binning of data, including an unbinned analysis. We provide more general explorations of this test statistic in Appendix \ref{sec:Stats}.

A particularly simple limit of \eqref{eq:LMod} is one for which we take two bins per day and describe the modulation simply by a single parameter, the integrated modulation fraction $f_2$, defined as the fractional difference in integrated rate between the two bins, averaged over a day: 
\begin{equation} \label{eq:frac-def}
    f_2 = \frac{1}{(24\, \text{h}) \langle R \rangle} \left( \int_{t_0}^{t_0+12{\rm h}} \! dt\, R(t) -  \int_{t_0+12{\rm h}}^{t_0+24{\rm h}} \! dt \, R(t) \right) .
\end{equation}
For a perfectly sinusoidal signal, $f_2$ equals the peak-to-trough amplitude divided by $\pi$. Our choice in \eqref{eq:vE} to align the crystalline symmetry axis, $\hat{b}$, with the lab frame DM wind at $t=0$ ensures that the dominant part of the modulation signal has a 24-hour period, with only small contributions from higher harmonics. In this orientation, the integrated modulation amplitude \eqref{eq:frac-def} is maximized by $t_0 \approx 18\,\text{hours}$, based on the results shown in Fig.~\ref{fig:rateNorm}. This observable is particularly well suited for describing the daily modulation, because it is unaffected by the non-modulating background rate and thus does not require any knowledge of the background.

As explored in detail in Appendix~\ref{sec:Stats}, this simple binning is amenable to analytic results in the large-$N$ limit of the Skellam distribution or in the small-modulation limit of the Poisson distribution. In each case, we find that the statistical significance we may assign to either the modulating or non-modulating hypothesis based on an experiment in which $N_{\rm tot}$ events are observed is
\begin{equation}
    N_\sigma= \frac{f_2\,  T_\text{exp} \langle R \rangle }{\sqrt{ N_\text{tot}}} ,
    \label{eq:NSigma2Bin}
\end{equation}
where $\langle R \rangle$ is the time average of the signal event rate $R(t)$ defined in \eqref{eq:rangle}, and where $T_\text{exp}$ is the total exposure time for the experiment. Since the number of signal and background events both grow linearly with exposure, the significance of a modulating signal improves with exposure as long as the integrated modulation fraction $f_2$ is nonzero. Our \eqref{eq:NSigma2Bin} matches the $\chi^2_{sb}$ statistic suggested by Ref.~\cite{Geilhufe:2019ndy}.

\begin{figure*}
    \centering
    \includegraphics[width=0.98\textwidth]{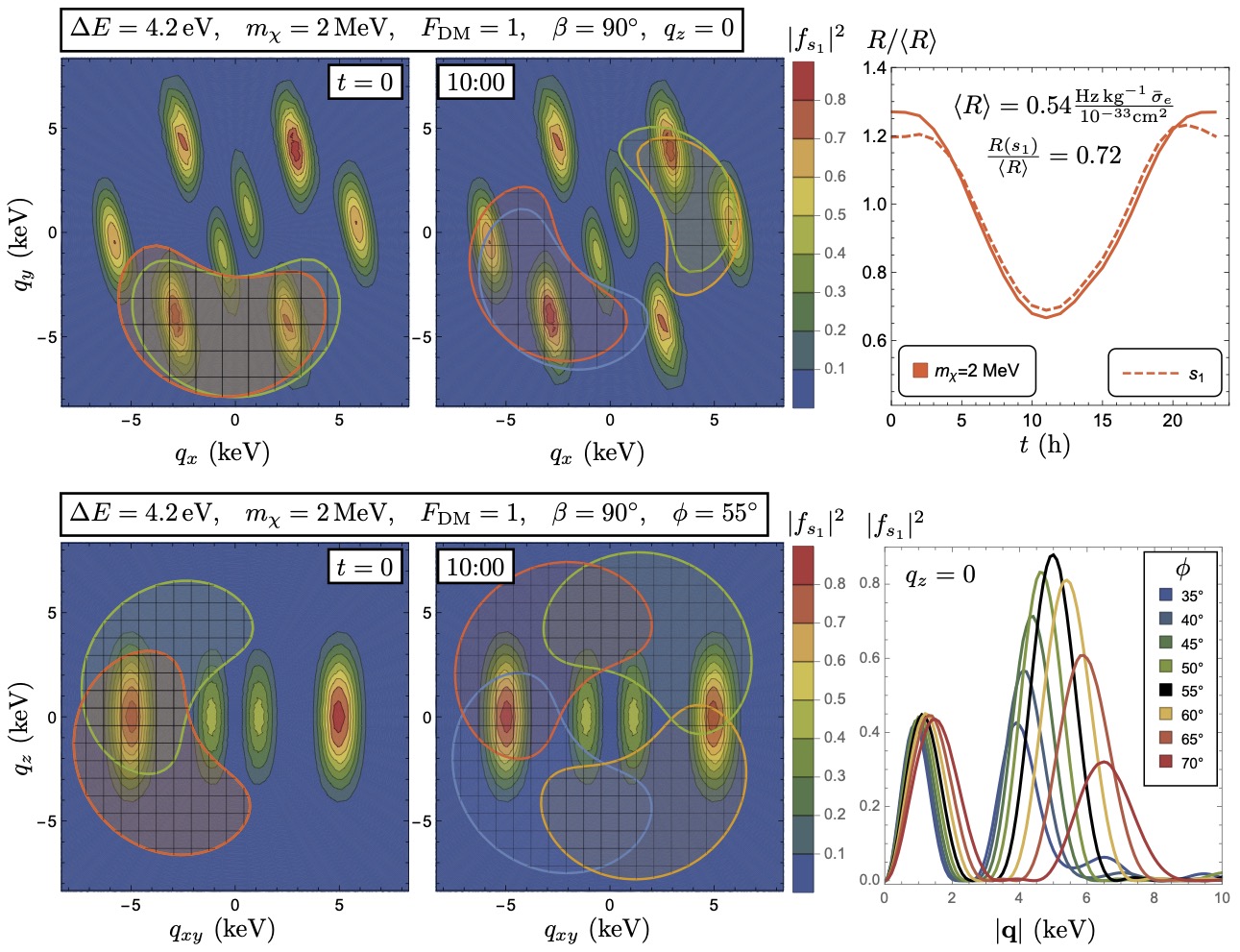}
    \caption{Molecular form factors and modulating rates for DM masses near threshold, $m_\chi = 2 \ {\rm MeV}$.  In the contour plots, the gridded shaded regions indicate the kinematically accessible momentum transfers $\vec{q}$ for the four molecules $M_{1, 2, -1, -2}$ that comprise the unit cell of the crystal, shown at $t=0$ and $t=10\,\text{h}$. Here, $\vec{q}$ is given in the molecular basis, $q_x = \vec{q} \cdot \hat{\vec{L}}$, $q_y = \vec{q} \cdot \hat{\vec{M}}$, and the kinematically accessible region is defined by $v_-(\vec{q}) < v_\text{esc}$, following \eqref{eq:vMinus}. \textbf{Top left:} Contour plot of the molecular form factor $|f(\vec{q})|^2$ for the $g \rightarrow s_1$ transition in the $(q_x, q_y)$ plane, with $q_z =0$. \textbf{Bottom left:} Contour plot for fixed $\phi\equiv \arctan (q_y/q_x) = 55^\circ$, showing the strong anisotropy in $q_z$ with maxima at $q_z = 0$. \textbf{Top right:} The scattering rate (summed over all $g \to s_i$ transitions) as a function of time, $R(t)$, normalized by the average daily rate $R_\text{avg}$. The modulation is dominated by the $s_1$ transition (dashed). \textbf{Bottom right:} A closer look at the form factor near the peak at $\phi \simeq 55^\circ$, plotting $|f_{s_1}|^2$ as a function of $|\vec{q}|$ for fixed $\theta=90^\circ$ and various $\phi$.
    }
    \label{fig:beanplots}
\end{figure*}

In Fig.~\ref{fig:exclusionsSHM} we show the expected results of a $1\, \text{kg}\cdot \text{year}$ t-stilbene experiment operated under a number of different assumptions. As a benchmark to facilitate comparison with other experiments, we demonstrate the reach with an entirely background-free experiment using no modulation information. The potential for parameter space exclusion in this scenario is $\bar \sigma_e \simeq 10^{-41} \cm^2 \ ({\rm few} \times 10^{-41} \cm^2)$ for DM interacting with a form factor $F_{\rm DM} = 1 \ (F_{\rm DM} \propto 1/q^2)$ and with a mass in the range $5 \MeV \lesssim m_\chi \lesssim 10 \MeV$. This is within a factor of 2 or 3 from the expected reach of a silicon CCD experiment like SENSEI or Oscura for an equivalent target mass \cite{Essig:2015cda,OSCURA}. Taking the more realistic scenario that the observed rate for a $1\,\text{kg}$ detector is $R=1\min^{-1} = 1/60\, {\rm Hz}$ (including both signal and background components), the future reach depends on analysis strategy. Without leveraging modulation information, the limit we obtain is slightly stronger than the current exclusion from SENSEI~\cite{Barak:2020fql} at low masses below $\sim 5 \MeV$, and comparable at higher masses. We also comment in passing on the prospects for the detectability of the (non-modulating) absorption of DM: since $\rho_T \simeq 1\, {\rm g/cm^3}$, we anticipate a rate of $\sim \mathcal O (1)/{\rm kg/min}$ for a dark photon kinetic mixing parameter of $\epsilon \simeq 10^{-13}$, assuming a dielectric loss of order $\sim \mathcal O(10^{-2})$, similar to that in benzene \cite{ronne2000benzenedielectric} and comparable to those in semiconductors \cite{Hochberg:2016sqx}. This setup would set leading limits on dark photons in the mass range $4.2 \eV < m_{A'} \lesssim 10 \eV$.

For DM scattering, our sensitivity to exclusion and discovery can be dramatically extended by utilizing the information in the rate via the simple two-bin analysis. Using the significance from \eqref{eq:NSigma2Bin}, a $1\, \text{kg}$, $1\,\text{year}$ t-stilbene experiment that observes a constant $R=1\min^{-1} = 1/60\, {\rm Hz}$ event rate can exclude at 90\% CL a DM particle with a scattering cross section as small as $\bar \sigma_e \simeq 10^{-37} \cm^2$. This cross section lies below the well-motivated line from freeze-out production of scalar DM for $2\MeV \lesssim m_\chi \lesssim 7\MeV$ with a heavy dark photon mediator mediator, and also probes a wide range of masses $2\MeV \lesssim m_\chi \lesssim 200\MeV$ for freeze-in production through a light mediator~\cite{Battaglieri:2017aum}. The $3\sigma$ discovery reach for a modulating signal for a total $R= 1/60\, {\rm Hz}/\text{kg}$ background event rate is nearly as strong, reaching just below (above) the cross section $\bar \sigma_e \simeq 10^{-37} \cm^2$ for $F_{\rm DM} = 1 \  (F_{\rm DM} \propto 1/q^2)$.

Very meaningfully, as shown in \eqref{eq:NSigma2Bin}, the discovery or exclusion significance grows with cumulative exposure, even without background mitigation: this improvement in significance is absent in a non-modulating signal. We demonstrate this explicitly in Fig.~\ref{fig:discovery}, displaying the $90\%$ CL exclusion and the $3\sigma$ discovery reach for a t-stilbene experiment with a constant observed rate $R=1\min^{-1}{\rm kg}^{-1}$ and increasing exposures of $0.01, 1, 100$ kg$\cdot$yr. For the lowest exposure proposed here, the background rate is very nearly equal to the $2e^-$ rate observed by the SENSEI experiment with a $\sim$2g detector \cite{Barak:2020fql}. The sensitivity improves with $\sqrt{N_{\rm tot}}$, so given the assumption of constant total rate in counts per unit time per unit mass, the sensitivity improves with $\sqrt{\rm exposure}$. This conservative expectation for scaling of the background rate essentially assumes that bulk events will dominate the background. There will be an irreducible background from the low-energy tail of $^{14}{\rm C}$ decays which would yield only a single scintillation photon, but assuming scintillators can be manufactured with the $10^{-18}$ g/g $^{14}{\rm C}$ levels achieved by Borexino \cite{Arpesella:2001iz,Benziger:2007aq}, the \emph{total} $^{14}{\rm C}$ decay rate would be 0.01 events/min/kg, well below the background rates we have assumed here. These background rates also include radio contamination from heavy metals (e.g. Th and U) whose beta decay spectrum is not compatible with the one-photon signal that such an experiment will look for. Finally, the cosmic ray background is expected to be the bulk of the exogenous rate and the only major background that might vary over the time scale of a day. This rate can be minimized by running under sufficient overburden, and the daily modulation of this background is constrained to be $\lesssim$ 0.1\% in a deep underground facility~\cite{COSINE:2020jml}. If the background rate were dominated by the dark rate in the photodetector, which would likely scale with area, a large-volume experiment and/or light-focusing scheme would improve the significance even further. We plan to return to these issues in future work.

\section{Kinematics and Target Selection For Daily Modulation}
\label{sec:Modulation}

\begin{figure*}
    \centering
    \includegraphics[width=0.98\textwidth]{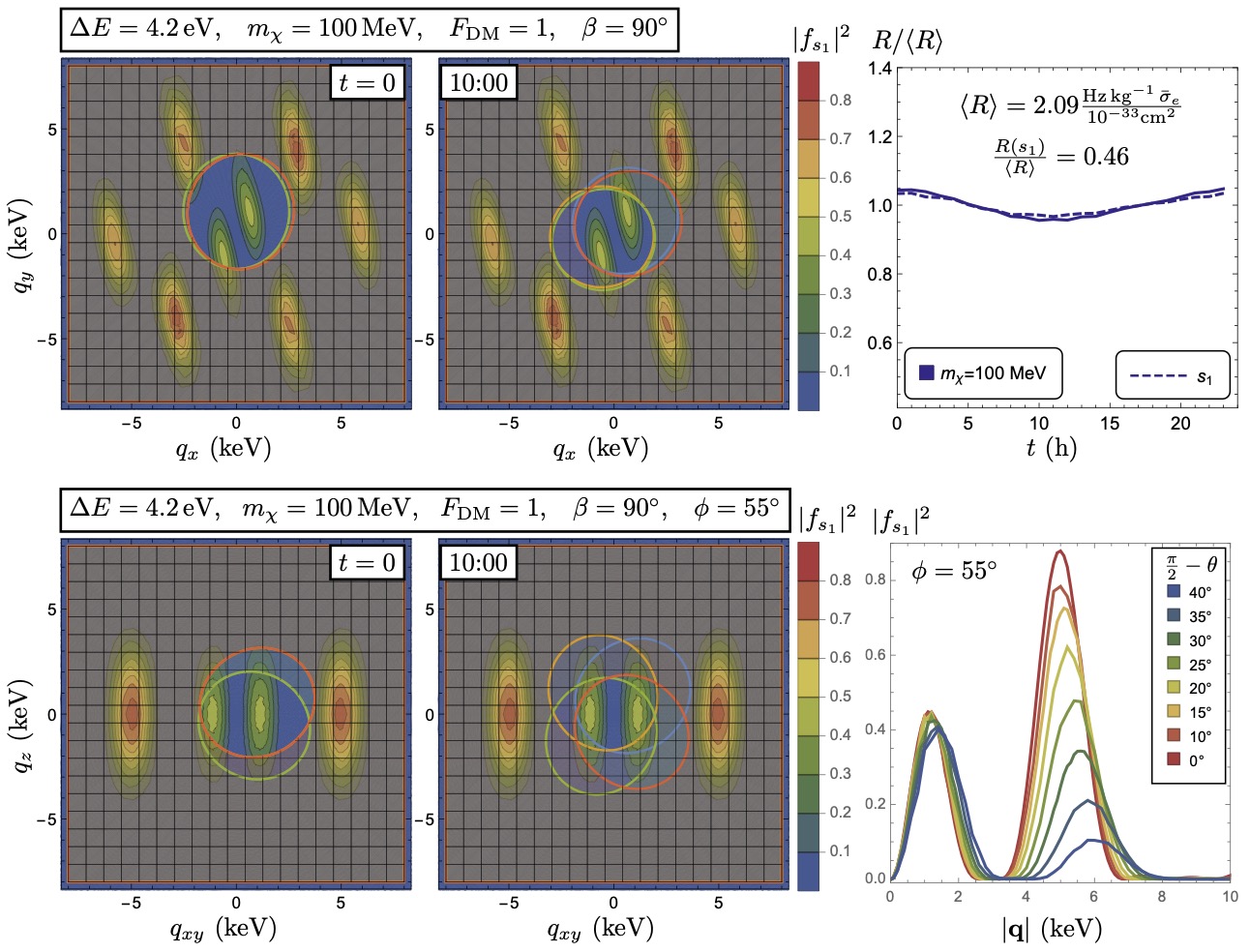}
    \caption{Same as Fig.~\ref{fig:beanplots} for large DM masses, $m_\chi = 100 \ {\rm MeV}$. Only the nearly-spherical region near $q \sim 0$ with inner boundary $q_{\rm min} \simeq 1.6 \ {\rm keV}$ is kinematically forbidden. As a result, the daily modulation amplitude is smaller, driven by the anisotropy of the inner secondary peaks and the tails of the primary peaks. 
    }
    \label{fig:beanplots100MeV}
\end{figure*}

\begin{figure*}
    \centering
    \includegraphics[width=0.98\textwidth]{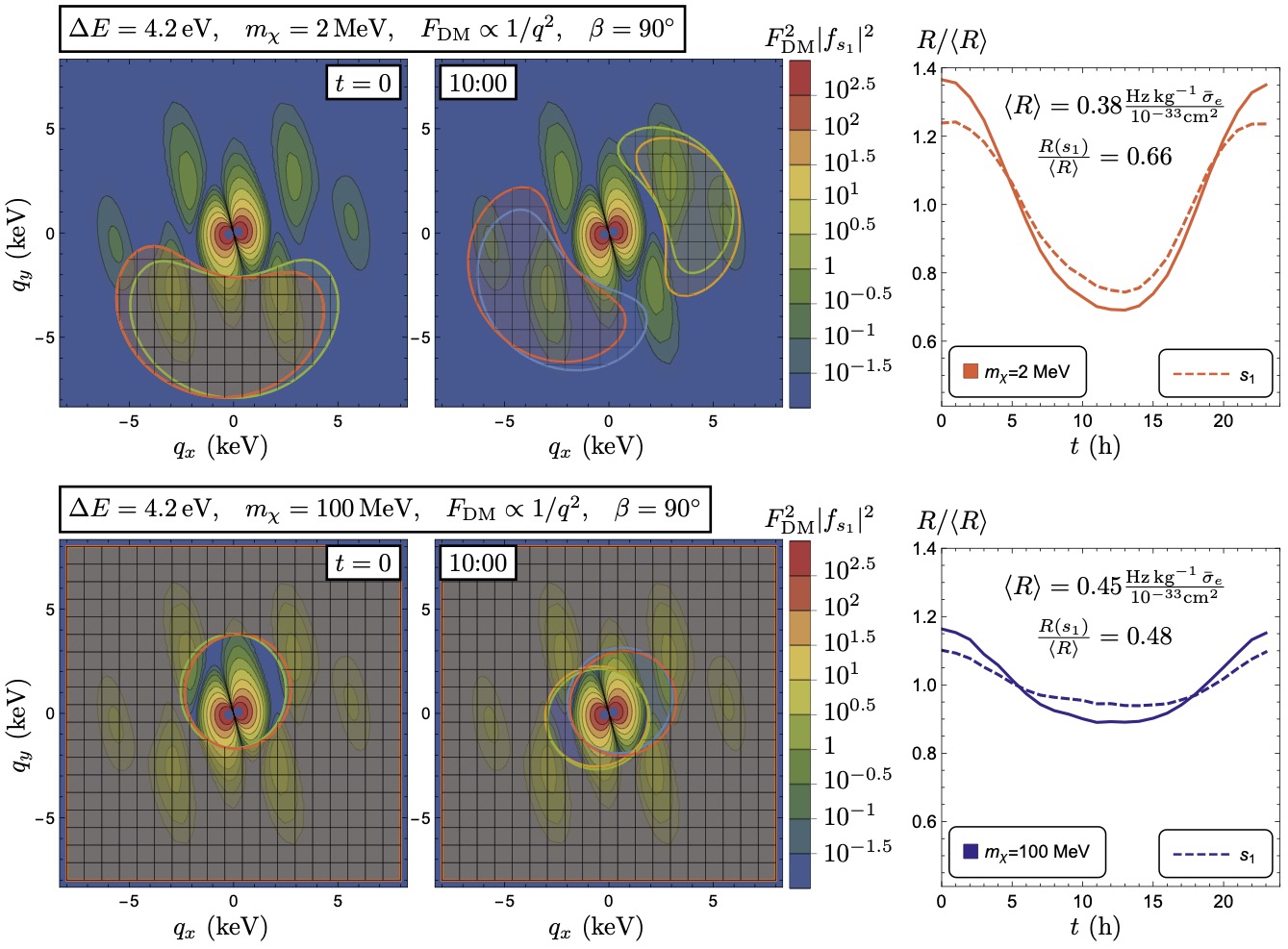}
    \caption{ Same as Figs.~\ref{fig:beanplots} and~\ref{fig:beanplots100MeV} (top) for a light mediator DM form factor $F_\text{DM} =(\alpha m_e/q)^2$. Here, the contour plots show $F_\text{DM}^2 |f(s_1)|^2$ which appears in the rate integrand \eqref{eq:RTotg0}; the scattering is dominated by the smallest kinematically-allowed $q$. \textbf{Top:} Molecular form factors with $q_z = 0$ and rate modulations for $m_\chi = 2 \ {\rm MeV}$.  \textbf{Bottom:} Molecular form factors with $q_z = 0$ and rate modulations for $m_\chi = 100 \ {\rm MeV}$. 
    }
    \label{fig:beanplotsFq2}
\end{figure*}

\begin{figure*}[t]
    \centering
    \includegraphics[width=0.98\textwidth]{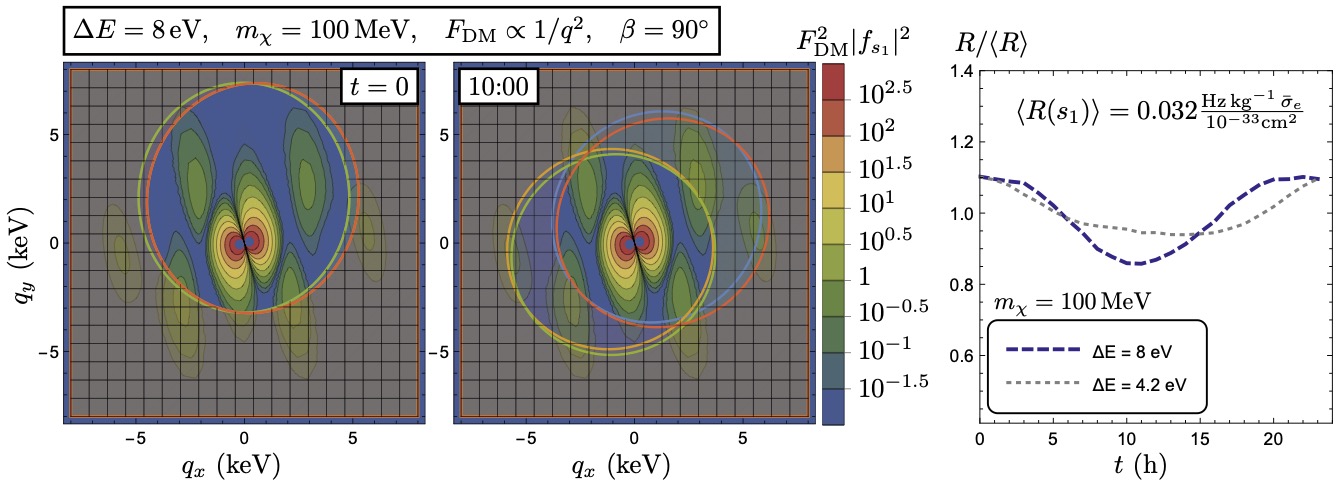}
    \caption{Same as Fig.~\ref{fig:beanplotsFq2} (bottom), but with a counterfactual transition energy of $\Delta E = 8 \ {\rm eV}$. This larger transition energy leads to a larger daily modulation for the $g \to s_1$ transition in the large $m_\chi$ limit. To facilitate a comparison with Fig.~\ref{fig:beanplots100MeV}, we also provide the modulation signal of the $s_1$ transition with $\Delta E = 4.2\, \eV$ as the dotted gray line with the smaller amplitude.
    }
    \label{fig:counterfactuals}
\end{figure*}

Given the large daily modulation amplitude present in t-stilbene, and the associated improvement in discovery and exclusion significance, it is worth examining which characteristics of our target molecule govern the size of the modulation, and whether other choices of organic molecules could improve the modulation amplitude even further. Indeed, in other systems sensitive to sub-MeV DM (Dirac materials, for example), the modulation amplitude can be even larger, $\mathcal{O}(1)$ even for DM masses well above threshold \cite{Geilhufe:2019ndy,Coskuner:2019odd}.

\subsection{Daily modulation in t-stilbene}
\label{sec:StilbeneKinematics}
The peaks of the t-stilbene molecular form factors define a preferred momentum scale $q^* \simeq 6 \ {\rm keV}$ where the rate is largest, so for the purposes of understanding the daily modulation, we may approximate all DM interactions as imparting momentum $q^*$. The $s_1$ transition has $\Delta E = 4.2 \ {\rm eV}$, which defines an effective velocity scale
\begin{equation}
    v^* \equiv \frac{\Delta E}{q^*} \simeq 200 \ {\rm km/s},
\end{equation}
on the same order as $v_\oplus \simeq 230 \ {\rm km/s}$. For sufficiently small $m_\chi$ such that $q^*/(2m_\chi) \simeq v_{\rm esc}$, Eq.~(\ref{eq:vMinus}) shows that $v_-(\vec{q})$ will be driven to $v_{\rm esc}$ unless $\vec{v}_\oplus$ is antiparallel to $\vec{q}$, and hence the rate will be nonzero only for a very narrow range of directions of $\vec{q}$. 

Fig.~\ref{fig:beanplots} illustrates this phenomenon, with the gridded ``bean-shaped'' shaded regions representing the kinematically-accessible region $v_-(\vec{q}) < v_\text{esc}$ overlaid on contour plots of the $s_1$ molecular form factor at $\beta = 90^\circ$ for DM mass of 2 MeV and a heavy-mediator form factor $F_{\rm DM} = 1$. There are four such regions for the four different t-stilbene orientations within a unit cell. The daily modulation arises from the movement of the kinematically-allowed region over the course of a day, in particular as these regions rotate out of the plane of the molecule and the peaks at $q_z = 0$ become inaccessible.

On the other hand, for sufficiently large $m_\chi$, $q/(2m_\chi) \to 0$ and $v_-(\vec{q^*}) < v_{\rm esc}$ for any direction of $\hat{\vec{q}}$. Thus, the kinematically-allowed region in $\vec{q}$-space always includes $q^*$ but has inner boundary
\begin{equation}
q_{\rm min} = \frac{\Delta E}{v_{\rm esc} + v_\oplus} \simeq 1.6 \ {\rm keV}.
\end{equation}
Fig.~\ref{fig:beanplots100MeV} provides the same information as Fig.~\ref{fig:beanplots} except now for a heavier DM particle, with mass $m_\chi = 100\MeV$. The form factor remains the same, but the ``beans'' have now expanded to fill in across the plane, leaving only circular ``holes'' with inner boundary $q_{\rm min}$. Because the kinematically-accessible region now includes the full peaks of the form factor, the rate modulation of the course of the day arises only due to the mismatch of the circular inner boundary with the hexagonal symmetry of the form factor and the presence of the inner secondary peaks, compounded by the vector addition of $\hat q$ and $\hat v_\oplus$. This leads to a smaller $\sim 10$\% peak-to-trough modulation amplitude for all $m_\chi \gtrsim 10 \ {\rm MeV}$.

For DM scattering through a light mediator, $F_{\rm DM}=(\alpha m_e/q)^2$, the rate integrand \eqref{eq:RTotg0} is weighted toward small $q$. In Fig.~\ref{fig:beanplotsFq2} we show the molecular form factors multiplied by $F_{\rm DM}^2$; the rescaled form factors are peaked more strongly towards low momenta, as expected. Because the inner peaks are kinematically forbidden for DM of all masses, but the tails of these peaks are also probed by all DM masses, this increases the magnitude of the peak-to-trough modulation amplitude to $ \simeq 70\%$ for $m_\chi = 2 \MeV$ and remains as large as $\simeq 30\%$ for $m_\chi = 100\MeV$.

\subsection{Target selection for daily modulation}

The analysis of Sec.~\ref{sec:StilbeneKinematics} suggests a strategy for designing target materials to obtain a large anisotropic response to electron scattering, and correspondingly large rate modulation, even in the limit of heavy DM. In this limit, the time-independent part of the argument of $v_-$ (\eqref{eq:vMinus}) is simply $\Delta E/q$. Now consider a material with a form factor peaked at a momentum $q^*$ and with a lowest-lying excitation energy $\Delta E$. To maximize the modulation, we look for a material for which $q^*$ and $\Delta E$ are related by $q^* \simeq \frac{\Delta E}{v_{\rm max}}$, where $v_{\rm max} =  v_{\rm esc} + v_\oplus$ is the maximum DM velocity attainable in the lab frame. Equivalently, the ``effective velocity'' characterizing the lowest-lying molecular transition, defined as $v^* = \Delta E/q^*$, should be $v^* \lesssim v_{\rm max} $. In t-stilbene, the primary outer peaks have $v^* \simeq 200 \ {\rm km/s}$, which is a factor of a few too small to lead to the maximal rate, whereas the secondary inner peaks have $v^* \simeq 1200 \ {\rm km/s}$, and these peaks are always kinematically forbidden. 

An ideal target for daily modulation would have either larger $\Delta E$ or a larger spatial extent (smaller $q^*$), so as to match $v^* \lesssim v_{\rm max}$ for the primary peaks. To illustrate this, Fig.~\ref{fig:counterfactuals} shows the molecular form factor for the $g \to s_1$ transition in t-stilbene but with the kinematically-allowed region defined by a transition energy $\Delta E = 8 \ {\rm eV}$, rather than the 4.2 eV in t-stilbene. Here, we have chosen a form factor $F_{\rm DM} \propto 1/q^2$, which weights the kinematically-forbidden inner peaks more, but the modulation is still driven by the forbidden region in $q$ which has comparable radius to the outer peaks. As the forbidden region moves in $\vec{q}$-space, the peak-to-trough modulation amplitude can be as large as 20\% for all $m_\chi \gtrsim 20 \ {\rm MeV}$ for the $g \to s_1$ transition alone in this hypothetical material, almost a factor of 2 larger than the modulation amplitude for the equivalent transition in t-stilbene.

Taking a broader perspective, the anisotropic response of a condensed matter target to DM-electron scattering arises from an interplay of preferred scales $q^*$ set by the molecular size and a coincidence between the effective transition velocity $v^*$ and the maximum DM velocity in the lab frame. In the case of organic molecular solids, the conjugated $\pi$-electron system provides two length (momentum) scales given by the extent of the molecule along the molecular plane ($q^* \simeq 1.2 \ {\rm keV}$ for t-stilbene), and the extent of a single 2p orbital $(q^* \simeq 6 \ {\rm keV})$, which sets both the carbon-carbon bond length and the extent of the out-of-plane $\pi$ orbitals. The large hierarchy between these scales in large organic molecules means that the excitation dynamics in the plane are largely separated from excitations along the normal to this plane. Transitions along the normal direction will typically require larger imparted momenta than in the extended directions, and thus the form factor for the lowest transition will be peaked at $q_z = 0$, with peaks in the $x-y$ plane corresponding to the characteristic scales of the molecular (sub)structure. The kinematically-allowed regions which dominate the rate integral rotate in $\vec{q}$-space over the course of the day, where the planar anisotropy (and to a lesser extent, the hexagonal structure of a benzene ring which breaks rotational symmetry to a discrete subgroup) gives the modulation for small $m_\chi$, and the anisotropy of the two displaced benzene rings contributes significantly to the residual modulation for large $m_\chi$. Having electronic transitions with $v^*$ slightly smaller than $v_{\rm max}$ (as in our counterfactual example with $\Delta E = 8 \ {\rm eV}$ in Fig.~\ref{fig:counterfactuals}) will maximize the anisotropy for masses above threshold. That said, there is an inevitable tradeoff between the modulation amplitude and the total rate (consistent with the analysis of Ref.~\cite{Coskuner:2021qxo} for single-phonon production) because as $v^*$ approaches $v_{\rm max}$, the kinematically-allowed transitions rely more and more on the high-velocity tail of the DM velocity distribution.

From this perspective, we can understand why daily modulation amplitudes are typically small or nonexistent for electron scattering in conventional semiconductor and noble liquid detectors. In noble liquids, the filled electron shells are spherically symmetric (ignoring small effects due to van der Waals attraction and dimerization between noble atoms), and thus the form factor will be isotropic and no daily modulation will occur. On the other hand, solid-state lattices have only discrete translational symmetries, which may be expected to lead to anisotropies like those due to the hexagonal structure of the benzene rings. However, the dominant low-energy electronic transitions in conventional semiconductors with eV-scale gaps are due to \emph{delocalized valence electrons}, which lead to a continuous energy spectrum and smooth form factors without a preferred momentum scale, at least for $q$ smaller than the inverse lattice spacing $\sim 3 \ {\rm keV}$.\footnote{In fact, the form factors in silicon and germanium have peaks at $q = 0$ and $\Delta E = 18 \ {\rm eV}$ from the plasmon, as well as at $v^* = v_F \simeq 10^{-2}$ from the approximately free Fermi gas behavior of the valence electrons, but neither of these peaks are kinematically-accessible for halo DM \cite{Hochberg:2021pkt}.} For larger $q$, scattering will probe core electron shells of single atoms at individual lattice sites, but these filled shells will be spherically symmetric and give isotropic form factors.  That said, more exotic solid-state systems like Dirac materials, where a combination of a narrow gap (which permits small $q$) and an anisotropic linear dispersion $\Delta E \sim \sqrt{v_x^2 q_x^2 + v_y^2 q_y^2 + v_z^2 q_z^2}$ with the $v_i$ bracketing $v_{\rm max}$, can have order-1 daily modulation \cite{Coskuner:2019odd,Geilhufe:2019ndy} and a fairly large overall rate \cite{Hochberg:2017wce}.

Importantly, $q^*$ is related to the characteristic size of the (sub)structure of the molecule as well as the symmetry of the transition, which determines whether the transition is dipole/quadrupole allowed. Meanwhile, the minimum $\Delta E$ is set by the HOMO/LUMO gap which is sensitive to the topology of the conjugated electron system, as well as the presence of functional groups which could donate or accept electronic density. For example, 1,2-diphenylacetylene is the acetylene-bridged analog of t-stilbene and presents a HOMO/LUMO gap of $\mathcal{O}(10\%)$ larger than that of t-stilbene~\cite{Takabe1974}. This implies that the two quantities are somewhat decoupled and can be independently tuned, at least for $\Delta E$ in the range $1-10$ eV where the efficiency for scintillation photon detection is high. Furthermore, theoretical computations of the DM form factor as detailed in this paper can be verified by the complementary experimental probe known as electron energy-loss spectroscopy (EELS), which can be used to extract the generalized oscillator strength (\textit{i.e.}~dielectric function) of the molecular excitations in a given target; such a measurement automatically includes many-body effects~\cite{Hochberg:2021pkt} such as the ones parameterized by the PPP Hamiltonian as well as multi-electron excitations. Since t-stilbene and related chromophores have been identified as good scintillators for decades, their single-crystal synthesis is mature and can be scaled up to $\mathcal{O}(10\; \text{cm})$ crystals~\cite{Carman_2013}. Therefore, it is entirely within the reach of existing methods and technology to implement $\mathcal{O}(10\; \text{kg})$ of organic crystal scintillation target since $\mathcal{O}(10\; \text{cm})$ single crystals of anthracene, t-stilbene and p-terphenyl are commercially available already, though ensuring the radiopurity of samples will be paramount to reduce backgrounds. In principle, this is no more challenging than obtaining radiopure liquid scintillator since crystals are readily grown from liquid stock.

\subsection{Daily modulation from dark matter kinematic substructure}

Another mechanism for exploring different scattering kinematics is supplied in the form of a cold, co-rotating stream of DM particles. The exemplary such sub-distribution of DM particles is the putative Nyx stream \cite{Necib:2019zbk}. This stream has velocity $\vec v_{\rm Nyx} \simeq (150, 0, 140)$ km/s \cite{Adhikari:2020gxw} for components $(v_r, v_\theta, v_\phi)$. In reality Nyx appears slightly anisotropic \cite{Necib:2019zbk, Adhikari:2020gxw}, but it has a relatively low spatially average velocity dispersion, $\bar \sigma_{\rm Nyx} \simeq 60$ km/s. Given its inferred size, we can attribute to it a low escape velocity $w_{\rm esc} = 150$ km/s. To calculate the rate for the Nyx stream, $\vec v_{\rm Nyx}$ is subtracted from $\vec v_\oplus \simeq (40, 10, 230)$ km/s \cite{OHare:2019qxc, Adhikari:2020gxw} in \eqref{eq:vMinus} and $\sigma_{\rm Nyx}$ and $w_{\rm esc}$ replace $v_0/\sqrt2$ and $v_{\rm esc}$ in \eqref{vel-g}, respectively. 

Because of the smaller escape velocity $w_\text{esc}$, activating the $4.2\,\eV$ transition requires larger momentum transfers,  $q \gtrsim \mathcal O(10)\,\keV$. As a result the inner peaks of $|f_{s_i}|^2$ at $q \simeq 1.2 \ \keV$ are kinematically inaccessible, and the peaks at $q^* \simeq 6 \ {\rm keV}$ are only accessible at the high velocity tail of the distribution $w \approx w_\text{esc}$, even for $m_\chi \gtrsim 100\,\MeV$. Keeping the peaks of $|f_{s_i}|^2$ at the edge of the kinematically-accessible region can induce a large modulation amplitude for a wide range of $m_\chi$, but at the price of significantly lowering the overall scattering rate. Because the Nyx fraction is $\lesssim 10\%$ \cite{OHare:2019qxc} of the local DM density, however, such modulation is unlikely to be a dramatic effect in any experiment, especially for the $F_\text{DM} \propto 1/q^2$ form factor.

\section{Conclusions}
\label{sec:Conclusions}

Among the many target materials proposed for DM-electron scattering, few have demonstrated the necessary anisotropic response to probe the daily modulation of DM, and none (to our knowledge) optimized for DM from the MeV to GeV scale. In this paper we have shown that organic crystals are a promising family of targets with excellent prospects for daily modulation, already at the same level as the expected annual modulation signal for the particular case of t-stilbene, and possibly larger if a compound with a suitable $v^* = \Delta E/q^*$ can be identified. In previous work \cite{Blanco:2019lrf} we have already demonstrated the efficacy of (liquid) organic scintillators in an experimental context, and we expect that many of the same design considerations will hold for solid-state scintillators.

The excellent overall sensitivity of t-stilbene -- within a factor of a few of a comparable mass of silicon -- combined with the additional handle of daily modulation, would make such a detector strongly complementary to the existing experimental program for Oscura \cite{OSCURA} which uses silicon targets. In the event a positive signal is detected, daily modulation will be crucial for confirming a DM origin, and we have also derived a useful test statistic for determining the daily modulation significance in the presence of non-modulating backgrounds. We will explore design considerations for a concrete experimental implementation of a crystal organic scintillator detector in future work.

Beyond the particular case of t-stilbene, we have argued that aromatic organic crystals are a near-optimal compromise between overall rate, daily modulation, and scalability to large target masses, for DM of mass $m_\chi \gtrsim 1 \ {\rm MeV}$. The building blocks of organic scintillators, the sp$^2$-hybridized carbon orbital and its 2p${}_z$ double bonding counterpart, are naturally anisotropic and support delocalized electronic states extended in the molecular plane, while the spatial extent of the 2p orbitals determines a preferred momentum scale. The weak intermolecular forces in organic crystals allow the electronic wavefunctions (and hence the form factors) to retain their molecular character rather than being entirely delocalized as in semiconductors. Furthermore, the discrete transitions at well-defined energies $\Delta E$, combined with the sharply-peaked form factors, give $v^*$ which is close to optimal for t-stilbene, and may give larger modulation in compounds with slightly larger HOMO/LUMO gaps and therefore with a slightly larger $v^*$. The combination of exciting features demonstrated by these results point to the ability to probe extremely well-motivated parameter space with plausible near-future technology, even in the presence of realistic but significant background rates. This indicates great potential for anisotropic organic scintillator detectors.

~\\~\\~\noindent
\textbf{Acknowledgments.} We thank Dan Baxter, Juan Collar, Mike Crisler, Patrick Draper, Adam Kline, Noah Kurinsky, Danielle Norcini, Dario Rodrigues, and Javier Tiffenberg for helpful conversations. YK and CB thank Danna Freedman and Stephen von Kugelgen for numerous organic chemistry discussions. The work of YK and BL was supported in part by DOE grant DE-SC0015655. The work of CB was supported in part by NASA through the NASA Hubble Fellowship Program grant HST-HF2-51451.001-A awarded by the Space Telescope Science Institute, which is operated by the Association of Universities for Research in Astronomy, Inc., for NASA, under contract NAS5-26555  as well as by the European Research Council under grant 742104. Fermilab is operated by Fermi Research Alliance, LLC under Contract No. DE-AC02-07CH11359 with the United States Department of Energy. This work made use of the Illinois Campus Cluster, a computing resource that is operated by the Illinois Campus Cluster Program (ICCP) in conjunction with the National Center for Supercomputing Applications (NCSA) and which is supported by funds from the University of Illinois at Urbana-Champaign.

\appendix

\begin{figure*}
\centering
\includegraphics[width=0.83\textwidth]{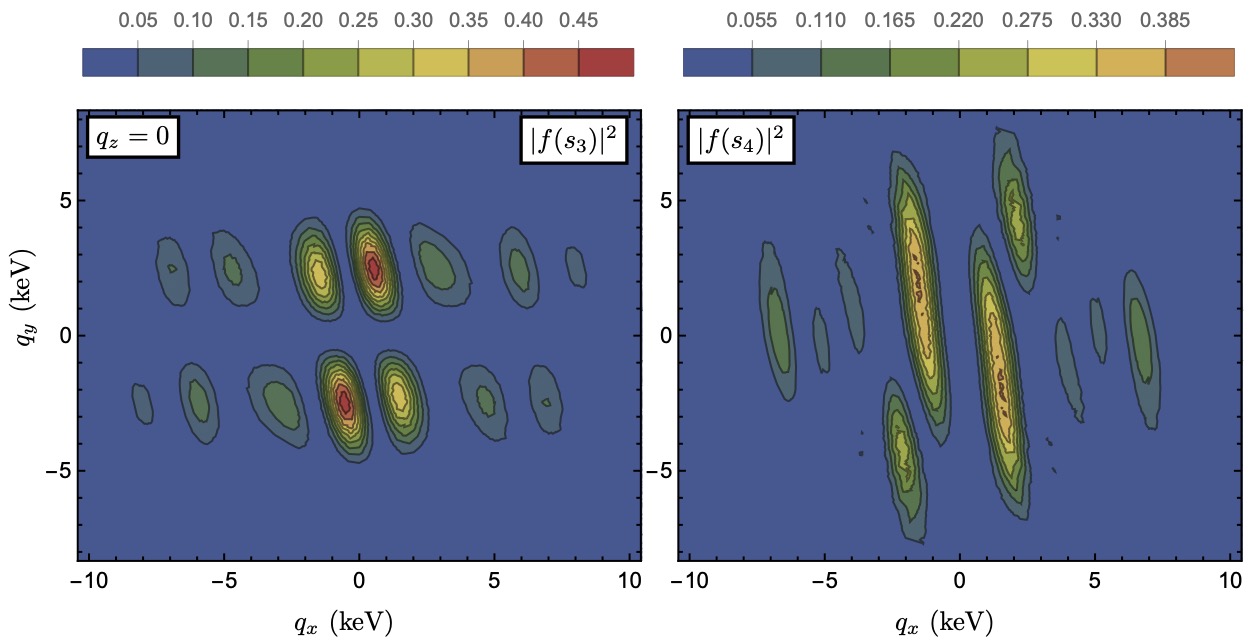}
\caption{Form factors $|f_{s_3}|^2$ and  $|f_{s_4}|^2$, shown in the $q_z = 0 $ plane as a function of $(q_x, q_y)$. The $s_4$ form factor has a roughly hexagonal structure, like the $s_1$ transition, but stretched in the $\pm q_y$ directions. The $s_3$ form factor has approximately rectangular symmetry, stretched along the long axis ($\hat x = \hat {\vec {L}}$) of the molecule. These two transitions provide the largest corrections to the scattering rate, but remain subdominant to the  $g \rightarrow s_1$ transition even in the large $m_\chi$ limit. 
\label{fig:s3s4ffs} }
\end{figure*}

\section{Molecular orbital calculation details}
\label{sec:OrbitalDetails}
\subsection{LCAO H\"uckel  Molecular Orbitals}
The HMOs of t-stilbene are found using the same technique as in Ref.~\cite{Blanco:2019lrf}. Here, the onsite energy is taken to be $\mathcal{E}_c=-6.7\text{eV}$ and nearest-neighbor resonance integral given by the following~\cite{Hawke2008, harrison2012electronic},
\begin{equation}
    H_{i j}=-0.63\frac{\hbar^{2}}{m_{e} x_{i j}^{2}}\delta_{i,i\pm 1} = -2.45\text{eV},
\end{equation}
where $x_{ij}$ is the distance between atoms. It should be noted, however, that these parameters become irrelevant after obtaining the HMOs when taking into account configurational interactions since new semi-empirical parameters are used for the single and many-body integrals as described by Pariser, Pople, and Parr~\cite{Pariser1953,Pariser1953a,pople1955electronic}.

\subsection{Form Factor Details} \label{sec:ffdetails}

Of the excited states beyond $s_1$, the $s_3$ and $s_4$ transitions provide the largest corrections to the scattering rate for a wide range of DM masses. In Fig.~\ref{fig:s3s4ffs}  we show their form factors squared in the $q_z=0$ plane. The $s_3$ form factor has a larger amplitude, and an approximately rectangular symmetry. Compared to the $s_1$ form factor, $|f_{s_3}|^2$ extends to larger values of $q_x$, parallel to the long axis of the molecule. The $s_4$ transition exhibits a rougher version of the approximate hexagonal symmetry of $|f_{s_1}|^2$, but with each peak stretched along the $q_y$ direction, and with enhancements to the peaks at $\phi \sim 120^\circ$ and $\phi \sim -60^\circ$. This is because the $s_4$ transition corresponds to one-electron excitations confined to the phenyl rings, so one should expect more spread-out support in $q$-space. The localized character of the ${}^1 \! A\to {}^1 \! H^+$ transition, in Platt notation, is discussed in detail by Beveridge and Jaff\'e (see e.g., Fig.~4 in~\cite{Beveridge1965}).

\begin{figure*}
    \centering
    \includegraphics[width=0.75\textwidth]{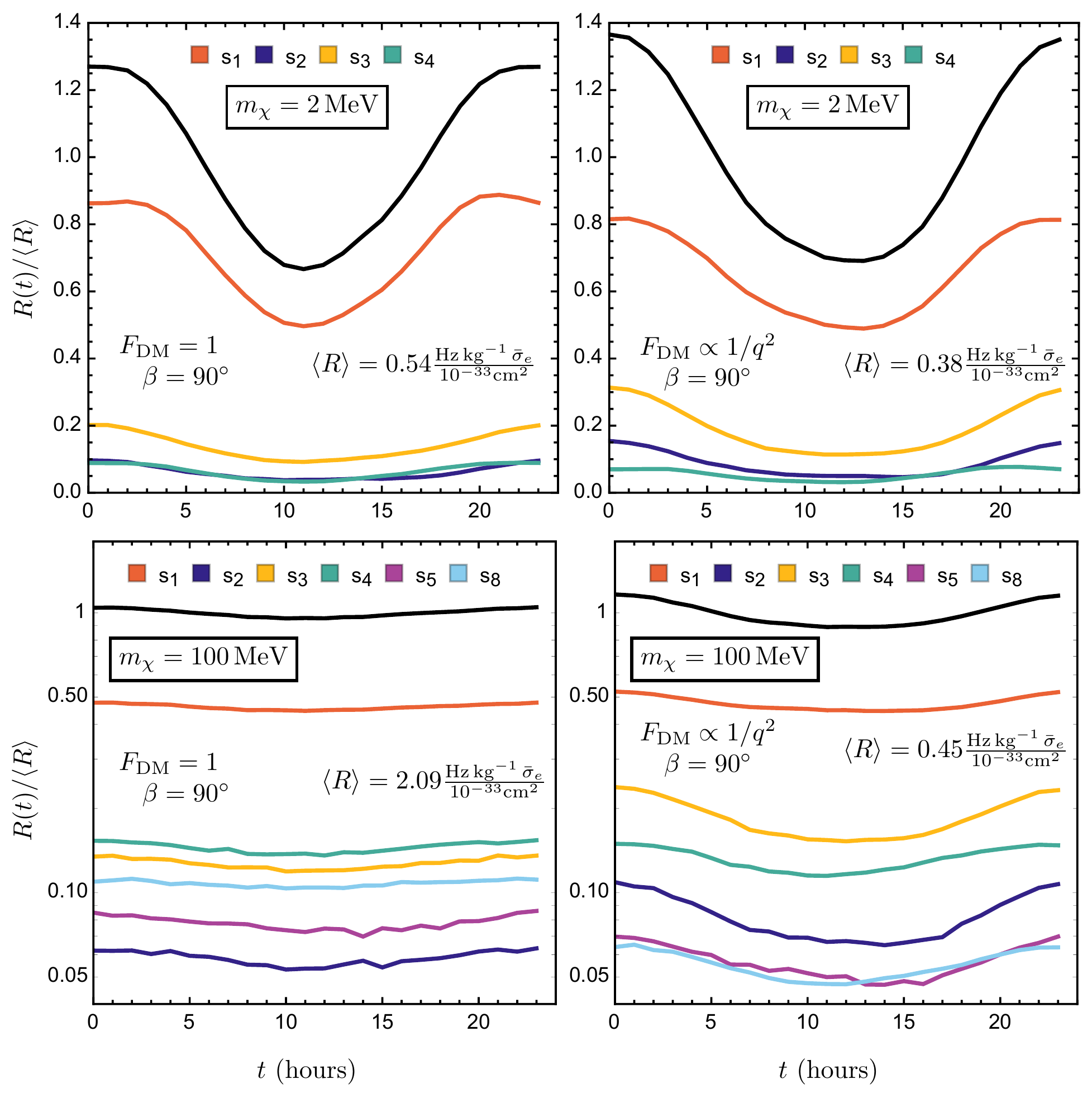}
    \caption{In the left and right panels we show the daily modulation signal from $2\,\MeV$ (top) and $100\,\MeV$ (bottom) DM in the $\beta=90^\circ$ orientation, with $F_\text{DM}=1$ and $F_\text{DM}=(\alpha m_e/q)^2$ in the left and right columns, respectively. The black lines show the total rate, normalized with respect to the 24~hour average, while each colored line represents the fraction of the signal that comes from an $s_{i}$ transition, also normalized by the total average rate. We use a logarithmic scale for the $100\,\MeV$ example, and include the $s_5$ and $s_8$ transitions. }
    \label{fig:rateByTransition}
\end{figure*}

The relative importance of each excited state to the scattering rate is shown in the fractional modulation plot of Fig.~\ref{fig:rateByTransition}. At $2\,\MeV$, around 65\%--70\% of the $F_{\rm DM}=1$ scattering rate occurs via the $g \rightarrow s_1$ transition. Around $15\%$ of the $F_{\rm DM}=1$ rate is due to the $s_3$ transition, with $s_2$ and $s_4$ each contributing at the 7--8\% level. For $F_\text{DM}=(\alpha m_e/q)^2$ the $s_3$ and $s_2$ transitions have somewhat greater importance, comprising $23\%$ and 11\% of the rate, respectively, compared to the 60\% fraction of the rate generated by $s_1$. We note that the conspicuously small contribution of $s_2$ serves as an independent confirmation of our molecular orbital model since the 1$^{1}A_{g}\to$2$^{1}B_{u}$ transition of t-stilbene in known to present an anomalously weak oscillator strength~\cite{Angeli2009}.

As indicated in Fig.~\ref{fig:beanplots100MeV} and Fig.~\ref{fig:beanplotsFq2}, the dominance of the $s_1$ transition is lessened at larger $m_\chi \gg 10\,\MeV$, due in part to the greater kinematic accessibility of the $s_5$ and $s_8$ transitions, but the $s_1$ still comprises nearly 50\% of the total scattering rate, with $s_3$ and $s_4$ providing the leading corrections. Furthermore, is it expected that at these larger $m_\chi$, $s_5$ and $s_8$ contribute a larger portion of the overall rate since they are the lowest lying  classically-allowed $^1B_u$ transitions with strong oscillator strengths. Meanwhile the hierarchy of the $s_3$ and $s_4$ transitions comes primarily from their $q$-dependent morphology since they are both $^1A_g$ transitions with roughly the same $\Delta E$.

\section{Statistics of daily modulation}
\label{sec:Stats}

Here we discuss two methods for obtaining analytic limits of \eqref{eq:LMod} and \eqref{eq:NSigma2Bin} in the simple two-bin analysis that we utilize in this paper.

\subsection{Skellam Distribution}\label{sec:skellam}
For a two-bin example, the likelihood and test statistic can be derived directly from the Skellam distribution~\cite{skellam1937}. Given two bins (1) and (2), with mean expected numbers of events $\mu_1$ and $\mu_2$, the probability of measuring a difference $\Delta N = N_1 - N_2$ between the numbers of events $N_{1,2}$ in the two bins is given by
\begin{align} \label{eq:skellam-def}
    &P\left( \Delta  N | \mu_1, \mu_2 \right) 
    = e^{-(\mu_1 + \mu_2)} \left( \frac{\mu_1}{\mu_2} \right)^{\Delta N /2} I_{\Delta N}\left(2\sqrt{\mu_1 \mu_2}\right),
\end{align}
where $I_k (z)$ is the $k$th modified Bessel function of the first kind, and where we have assumed Poisson statistics for the distribution of events in each bin.
Defining
\begin{equation}
    \mu_{\rm tot} \equiv \mu_1 + \mu_2, \qquad \mu_\Delta \equiv \mu_1 - \mu_2
\end{equation}
for convenience, the mean $\mu_0$, variance $\sigma^2$, skew and excess kurtosis of the Skellam distribution are~\cite{Lillard:2019exp}
\begin{align}
\mu_0 &= \mu_\Delta ,
&
\sigma^2 &= \mu_{\rm tot},
\label{eq:SkellamMoments}
\\
\gamma_1 &= \frac{\mu_\Delta}{\mu_{\rm tot}^{3/2}} ,
&
\gamma_2 &= \frac{1}{\mu_{\rm tot}},
\end{align}
so that in the large $\mu_{\rm tot}$ limit the distribution is approximately Gaussian.

An exact version of the test statistic can be derived from the double-sided distribution,
\begin{align}
L^{(h)} = - 2 \ln \lambda^{(h)},
&&
\lambda^{(h)} = \sum_{|j| \geq |\Delta N|} P(j | \mu_{1}^{(h)}, \mu_2^{(h)} ) ,
\label{eq:deflambda}
\end{align}
where the index $(h)$ refers to the null or modulating hypotheses, $(0)$ or $(m)$. The difference between the test statistics,
\begin{align}
    \Delta L \equiv L^{(m)} - L^{(0)} = - 2 \ln \frac{\lambda^{(m)} }{\lambda^{(0)}} ,
\end{align}
quantifies the significance of a signal and obeys a $\chi^2$ distribution with two degrees of freedom.

If $\mu_{1,2} \gg 3$, the Skellam distribution is well described by the Gaussian
\begin{align}
    P(\Delta N | \mu_1, \mu_2) &\simeq \frac{1}{\sqrt{2\pi \sigma^2}} \exp\left(- \frac{(\Delta N - \mu_\Delta)^2 }{2 \sigma^2}  \right),
\end{align}
with $\sigma^2 = \mu_{\rm tot}$, where the higher moments $\gamma_{1,2}$ become negligible. In this limit $\lambda$ can also be approximated by an integral over a continuous variable,
\begin{align}
    1 - \lambda &= \int_{\mu_\Delta - |\Delta N|}^{\mu_\Delta + |\Delta N|} \! dn\, P(n | \mu_1, \mu_2) \nonumber\\
    &\simeq \erf\left( \frac{|\Delta N - \mu_\Delta|}{\sqrt{2 \mu_{\rm tot}} } \right).
\end{align}
The significance of a measured $\Delta N$ can be easily expressed in terms of a number of standard deviations $N_\sigma$ by inverting the error function:
\begin{align}
    N_\sigma = \sqrt{2} \erf^{-1} \left( 1 - \lambda \right) 
    &\simeq \frac{ | \Delta N - \mu_\Delta | }{\sqrt{\mu_{\rm tot}}} .
    \label{eq:NsigmaDeltaN}
\end{align}

As an example, we apply this result to a modulating signal, $\mu_1^{(m)} \neq \mu_2^{(m)}$, and the null hypothesis of a constant background rate, $\mu_1^{(0)} = \mu_2^{(0)}$, where $\mu_1$ and $\mu_2$ are the predicted numbers of events in two bins of equal integration time. They can be expressed in terms of the expected signal and background rates ($\mu_s$ and $\mu_b$, respectively) for each hypothesis: 
\begin{align}
    \mu_1^{(m)} = \mu_s (1 + f_2) + & \mu_b , \qquad  \mu_2^{(m)}  = \mu_s(1 - f_2) + \mu_b, \nonumber \\
    &
    \mu_1^{(0)} = \mu_2^{(0)} = \mu_b^0,
\end{align}
where $f_2$ is the integrated modulation fraction defined in \eqref{eq:frac-def}, and the significance of a measurement of the number of events in each bin $N_1$ and $N_2$ can be assessed using \eqref{eq:NsigmaDeltaN}. In terms of $R(t)$ from \eqref{eq:RTotg0}, $\mu_s$ is 
\begin{align}
    \mu_s = \frac{1}{2} T_\text{exp} \langle R \rangle ,
\end{align}
where $\langle R \rangle$ is $R(t)$ averaged over one sidereal day, and $T_\text{exp}$ is the total exposure time.

If the background rate were well understood, the total number of events ($N_\text{tot} = N_1 + N_2$) could be compared to the prediction from the null hypothesis, $N_{\rm tot}^{(0)} = 2 \mu_b^0 = \mu_{\rm tot}$ as a way to discover or exclude particular DM models. Even without knowledge of the background, a small value for $N_\text{tot}$ can still be used to rule out those models which predict significantly more events than the measured $N_\text{tot}$, but a larger $N_\text{tot}$ cannot be construed as a detection of DM without a better understanding of the background. However, the existence of a modulating signal provides an additional statistical handle on both discovery and exclusion.

Assuming that the background rate is unmodelled, a measurement of $N_\text{tot}$ supplies the best-fit values for $\mu_b^{(0)}$ and $\mu_b^{(m)}$ in the null and modulating hypotheses, through
\begin{align}
    \mu_b^{(0)} &= \frac{1}{2} N_\text{tot}.
    &
        \mu_s + \mu_b^{(m)} &= \frac{1}{2} N_\text{tot}.
\end{align}
All of the information about the signal, $\mu_s$, is extracted from the measured value of $\Delta N = N_+ - N_-$, which has expected value
\begin{align}
    \langle \Delta N \rangle = 2 f_2(m_\chi, \sigmabare) \, \mu_s(m_\chi, \sigmabare),
\end{align}
and we have explicitly specified that both the modulation fraction $f$ and the signal strength parameter $\mu_s$ depend on the DM mass $m_\chi$ and cross section $\sigmabare$.

In assessing the capabilities of a directional detection experiment in the presence of daily modulation, we ask two questions: which DM models predict modulation signals that are large enough to be detected by the experiment? And, if the experiment measures a null result, which DM models are ruled out?
The first question, which determines the discovery significance, can be posed in the context of ruling out the null hypothesis, where $\mu_\Delta^{(0)} = 0$:
\begin{align}
    N_\sigma^\text{disc.}(\Delta N) \simeq \frac{ |\Delta N| }{\sqrt{\mu_{\rm tot}^{(0)}}} = \frac{|\Delta N | }{\sqrt{ N_\text{tot} } },
\end{align}
in the Gaussian limit $N_\text{tot} \gg 3$. Models that satisfy $N_\sigma^\text{disc.}(\langle \Delta N \rangle) > 3$ or $N_\sigma^\text{disc.}(\langle \Delta N \rangle)>5$ for this central value, for example, are likely to generate a modulation signal strong enough to claim a detection at the $3\sigma$ or $5\sigma$ level, respectively. 
The ``Discovery'' regions in Fig.~\ref{fig:exclusionsSHM} and Fig.~\ref{fig:discovery} show $N_\sigma^\text{disc.}(2 f_2 \mu_s) \geq 3$ using this central value.

To set exclusion limits, we ask which models are ruled out by a null result, $\Delta N \approx 0$. In this case, we use $\mu_\Delta^{(m)} = 2 f_2 \mu_s$ in \eqref{eq:NsigmaDeltaN}:
\begin{align}
    N_\sigma^\text{excl.}(\Delta N) \simeq \frac{ | \Delta N - \mu_\Delta^{(m)} | }{\sqrt{2(\mu_s + \mu_b^{(m)})}} = \frac{ | \Delta N - 2f_2 \mu_s | }{\sqrt{N_\text{tot}} } .
\end{align}
The exclusion curves of Fig.~\ref{fig:exclusionsSHM} and Fig.~\ref{fig:discovery} show the models that would be excluded at the 90\% confidence level from a null result $\Delta N = 0$, using $N_\sigma^\text{excl.}(0) > 1.65$. 

In both examples above, we have taken $\Delta N$ to be equal to its expectation value under either the modulating hypothesis ($\Delta N = 2 \mu_s f_2$) or the null hypothesis ($\Delta N = 0$), which has allowed us to quantify the significance of a measurement without reference to the log-likelihood ratio: both discovery and exclusion limits are given by
\begin{equation}
    N_\sigma = \frac{2 f_2 \mu_s}{\sqrt{N_{\rm tot}}}.
    \label{eq:NsigmaAll}
\end{equation}
This is simply because the ``$p$ value'', $\lambda$, is equal to $1$ at the central value of the double sided distribution \eqref{eq:deflambda}, and so the likelihood ratio is trivial. Because $\mu_s$ is the expected counts per {\it half} of a day, we see that $2 \mu_s$ is the exposure times the rate expected in \eqref{eq:RTotg0}. Thus, \eqref{eq:NsigmaAll} exactly recovers \eqref{eq:NSigma2Bin}.

To assess a measurement away from the central value with more generality, it is better to use the log-likelihood $L$ defined exactly in \eqref{eq:deflambda} for each hypothesis. For small $\lambda$, the $N_\sigma(\lambda)$ defined in \eqref{eq:NsigmaDeltaN} is approximated by
\begin{align}
N_\sigma(L) \approx \left( L + \ln\frac{2}{\pi L} + \frac{\ln \frac{\pi}{2}}{L} + \mathcal O(1/L^2)  \right)^{1/2}.
\label{eq:Nsigma}
\end{align}
For $N_\sigma \gg 1$, \eqref{eq:Nsigma} can be inverted to give an expression for $L$ in terms of $N_\sigma$ and expanded as a series in $N_\sigma^{-2}$,
\begin{align}
    L \simeq N_\sigma^2 + \ln\left( \frac{ N_\sigma^2 \pi}{2} \right) - \ln\left( 1 - \frac{1}{N_\sigma^2} %+ \frac{3}{N_\sigma^4} 
    + \ldots \right) .
\end{align}
This expression is particularly useful in the Gaussian limit, where $N_\sigma$ is given by \eqref{eq:NsigmaDeltaN}. To leading order in large $N_\sigma$,
\begin{align}
    \Delta L(\Delta N) \approx \frac{(\Delta N - 2 f_2 \mu_s)^2 }{N_\text{tot}} - \frac{(\Delta N)^2 }{N_\text{tot}},
\end{align}
where the test statistic $\Delta L$ compares a specific modulation model with $f_2(m_\chi, \sigma_e) \mu_s (m_\chi, \sigma_e)$ to the null hypothesis, $\mu_b^{(0)} = \frac{1}{2} N_\text{tot}$.
At the central values of the two distributions, $\Delta N = 2 f_2 \mu_s$ and $\Delta N = 0$, the test statistic takes values of $\Delta L = \mp 4 f_2^2 \mu_s^2/N_\text{tot}$, respectively, and we recover exactly \eqref{eq:NsigmaAll}. 

For other values of $\Delta N$, still in the Gaussian limit ($\mu_s \gg 3$), the significance is found from the cumulative distribution function (CDF) of the $\chi^2_{2}$ distribution,
\begin{align}
    {\rm CDF}(\Delta L) \approx  1 - \gamma\left(1, \frac{1}{2} \Delta L \right),
    \label{eq:pgamma}
\end{align}
where $\gamma$ is the lower incomplete Euler gamma function, and
where $k=1$ for the simple two-bin analysis.
Generalizing to $k$ statistically independent pairs of bins, the combined test statistic
\begin{align}
    \Delta L = \sum_{i=1}^k \Delta L_i 
\end{align}
satisfies a $\chi^2$ distribution with $2 k$ degrees of freedom,
and its CDF is given by 
\begin{align}
    p \approx 1 - \frac{\gamma\left(k, \frac{1}{2} \Delta L \right) }{\Gamma(k)}.
    \label{eq:pgammak}
\end{align}
In the limit of very few events, $\mu_s \lesssim 3$, the CDF should be evaluated using the Skellam distribution instead, as it ceases to be approximately Gaussian for $\mu_{1,2} < 1$. However, as it is not possible to resolve an $O(10\%)$ modulation fraction with so few events, in this limit a more powerful constraint will come from using Poisson statistics on the total number of events.

\subsection{Alternate Derivation with Poisson Statistics}
The negative log-likelihood for a Poisson process is
\begin{equation} \label{poiss-ll}
    L \equiv -2\ln \lambda = 2 \sum_{k=1}^{N_{\rm bins}} \left[ \nu_k(\bm{\theta}) - n_k + n_k \ln(n_k/\nu_k(\bm{\theta})) \right],
\end{equation}
where $\nu_k$ is the expected number of events in a bin $k$, $n_k$ is the observed number in that bin, and $\bm{\theta}$ are $N_\theta$ parameters that determine $\nu_k$. The statistic $L$ follows a $\chi^2$ distribution of $N_{\rm bins}-N_\theta$ degrees of freedom \cite{Zyla:2020zbs}.

If we want to compare the hypothesis of a modulating signal versus the null hypothesis of a nonmodulating signal, we simply take the difference of their log-likelihoods. Since the total number of events in a day is fixed in the two scenarios, and only their distribution throughout the day is varying, we have
\begin{equation} \label{delta-L-mod}
    \Delta L = - 2 \sum_{k=1}^{N_{\rm bins}} n_k \ln \left[\frac{\nu_k^m(\bm{\theta}^m)}{\nu_k^0(\bm{\theta}^0)} \right],
\end{equation}
where $\nu_k^m$ is the number of expected events assuming a modulating signal and $\nu_k^0$ is the number of events assuming a constant rate over the course of a day, and we have chosen the sign such that $\Delta L < 0$ means that modulation is preferred. The distribution of values of $|\Delta L|$ follows a $\chi^2$ distribution with the number of degrees of freedom set by the difference between the number of parameters $\bm \theta^m$ and the number of parameters $\bm \theta^0$. \eqref{delta-L-mod} is an exact expression for the improvement in fit when allowing a modulating signal instead of a constant signal, appropriate for whichever event binning is most convenient. Because the total number of events is fixed, this is also the difference of the Kullback-Leibler divergences between these two hypotheses with the data.

\eqref{poiss-ll} and \eqref{delta-L-mod} are appropriate for any choice of data binning, and can even be used for an unbinned analysis. For simplicity, and to provide an alternative derivation of the results in Sec.~\ref{sec:skellam}, we calculate \eqref{delta-L-mod} for the specific choice of two 12-hour bins per day, here labeled by $+$ and $-$. The rates per bin are $\nu_\pm^m = \bar \nu^s(1\pm f_2) + \nu^b$ and  $\nu_\pm^0 = \bar \nu^s + \nu^b$, where $\nu^b$ is the expected background rate and $\bar \nu^s$ is the average expected signal rate per bin. 

Let us assume first that the true signal is not modulating: the number of observed counts in each bin in a given day is expected to be equal, such that $\langle n_{k +} \rangle = \langle n_{k -} \rangle = \bar \nu^s + \nu^b$. In this case, we have

\begin{align}
    \label{eq:DL-2bin}
    \Delta L & = -2 \sum_{d=1}^{N_{\rm days}} \sum_\pm (\bar \nu^s + \nu^b) \ln \left( 1 \pm \frac{ f_2 \bar \nu^s}{\bar \nu^s + \nu^b} \right)
   \nonumber \\ &= - 2 \sum_{d=1}^{N_{\rm days}} (\bar \nu^s + \nu^b) \ln \left[ 1- \left( \frac{ f_2 \bar \nu^s}{\bar \nu^s + \nu^b} \right)^2 \right] 
  \nonumber  \\& \simeq 2\sum_{d=1}^{N_{\rm days}}  \frac{ f_2^2  (\bar \nu^s)^2}{\bar \nu^s + \nu^b},
\end{align}
where in the second step we take the limit $f_2 \bar \nu^s/(\bar \nu^s + \nu^b) \ll 1$. We now define $N_{\rm tot} = \sum_d 2( \bar \nu^s + \nu^b) = 2N_{\rm days}( \bar \nu^s + \nu^b)$ to be the total number of events observed and, to make contact with the preceding section, we define $\mu_s = \sum_d \bar \nu^s = N_{\rm days} \bar \nu^s$ to be half of the total number of signal events expected to be observed over the entire experimental exposure. This gives
\begin{equation} \label{eq:poiss-DeltaL}
    \Delta L \simeq \frac{(2 f_2 \mu_s)^2}{N_{\rm tot}}    
\end{equation}
The observed significance of a signal is therefore $ \chi^2 = (2f_2 \mu_s)^2/N_{\rm tot}$.
Conversely, a limit at $N_\sigma$ significance on a modulating signal is possible when $N_{\rm tot}^{\rm excl.}\simeq (2f_2 \mu_s / N_\sigma)^2$. As in the preceding section of this Appendix, because $\mu_s$ is the expected counts per {\it half} of a day, the factor $2 \mu_s$ is the exposure times the rate expected in \eqref{eq:RTotg0}. Thus, \eqref{eq:poiss-DeltaL} exactly recovers \eqref{eq:NSigma2Bin}.

Assuming on the other hand that the signal {\it is} modulating, the number of observed counts is no longer expected to be the same in the two bins in a given day. Instead, the counts will be related by $\langle n_{k \pm} \rangle = \bar \nu^s(1 \pm f_2) + \nu^b$. In this case, we have
% \begin{widetext}
% \alg{  \label{eq:DL-2bin-mod}
%     \Delta L  &= - 2 \sum_{d=1}^{N_{\rm days}} \sum_\pm [\bar \nu^s(1 \pm f_2) + \nu^b ] \ln \left( 1 \pm \frac{ f_2 \bar \nu^s}{\bar \nu^s + \nu^b} \right)
%     \\& \simeq - 2 \sum_{d=1}^{N_{\rm days}} \sum_\pm [\bar \nu^s + \nu^b \pm f_2\bar \nu^s] \left[ \pm \frac{ f_2 \bar \nu^s}{\bar \nu^s + \nu^b} - \frac12 \left( \frac{ f_2 \bar \nu^s}{\bar \nu^s + \nu^b} \right)^2 \right]
%     \\& \simeq - 2\sum_{d=1}^{N_{\rm days}}  \frac{ f_2^2  (\bar \nu^s)^2}{\bar \nu^s + \nu^b},
% }
% \end{widetext}
\begin{widetext}
\begin{align}
  \label{eq:DL-2bin-mod}
    \Delta L  &= - 2 \sum_{d=1}^{N_{\rm days}} \sum_\pm [\bar \nu^s(1 \pm f_2) + \nu^b ] \ln \left( 1 \pm \frac{ f_2 \bar \nu^s}{\bar \nu^s + \nu^b} \right)
   \nonumber \\& \simeq - 2 \sum_{d=1}^{N_{\rm days}} \sum_\pm [\bar \nu^s + \nu^b \pm f_2\bar \nu^s] \left[ \pm \frac{ f_2 \bar \nu^s}{\bar \nu^s + \nu^b} - \frac12 \left( \frac{ f_2 \bar \nu^s}{\bar \nu^s + \nu^b} \right)^2 \right]
  \nonumber  \\& \simeq - 2\sum_{d=1}^{N_{\rm days}}  \frac{ f_2^2  (\bar \nu^s)^2}{\bar \nu^s + \nu^b},
\end{align}
\end{widetext}
where the relative sign between \eqref{eq:DL-2bin} and \eqref{eq:DL-2bin-mod} is reflective of the choice we made that $\Delta L < 0$ means that a modulating signal is preferred. The magnitude of the significance is exactly the same as in the prior case, differing only in that the interpretation in this scenario is as discovery of a signal.

\bibliography{OrganicScintillator.bib}

\end{document}